\documentclass[a4paper,12pt,preprint,nofootinbib]{revtex4}
\usepackage{amsmath,mathrsfs,amscd,graphicx}
\usepackage{mciteplus}
\usepackage{multirow}
\usepackage{color}
\usepackage{rotating}
\usepackage{slashed}
\usepackage{subfig}
\usepackage{textcomp}
\usepackage[final]{pdfpages}
\usepackage{array}
\usepackage{float}
\usepackage{url}
\usepackage{textcomp}
\usepackage{amsfonts, latexsym, epsfig}
\usepackage{bm}
\usepackage{times}
\usepackage{epsfig}
\usepackage{amssymb}
\usepackage{tikz}
\usepackage{cancel}
\usepackage{hyperref}
\usepackage{verbatim}
\usepackage[normalem]{ulem}

\def\beq{\begin{equation}}
\def\eeq{\end{equation}}
\def\be{\begin{equation}}
\def\ee{\end{equation}}
\def\bea{\begin{eqnarray}}
\def\eea{\end{eqnarray}}

\begin{document}

\title{Probe Charm Yukawa at the future $e^{-}p$ and $e^{+}e^{-}$ colliders}
\author{Ruibo Li~}
\email{bobli@zju.edu.cn}
\author{Bo-Wen Wang~}
\email{0617626@zju.edu.cn}
\author{Kai Wang~}
\email{wangkai1@zju.edu.cn}
\author{Xiaoyuan Zhang~}
\email{3160103951@zju.edu.cn}
\author{Zhenyu Zhou~}
\email{21736011@zju.edu.cn}
\affiliation{
 Zhejiang Institute of Modern Physics, Department of Physics, Zhejiang University, Hangzhou, Zhejiang 310027, CHINA
}

\begin{abstract}

Large Hadron Collider (LHC) has provided direct evidence of Yukawa couplings between the third generation charged fermions and the 125~GeV Higgs boson. Whether the first two generation charged fermions arise from exactly the same mechanism becomes the next interesting question. Therefore, direct measurements of charm or muon Yukawa couplings will be crucial to answering this puzzle. The charm Yukawa measurement at the LHC suffers from severe QCD background and it is extremely difficult to reach the sensitivity. In this paper, we compare the potential of probing charm Yukawa coupling at the two proposed future ``Higgs Factory" experiments, the Large Hadron electron Collider (LHeC) and Circular electron positron collider (CEPC). At the LHeC, Higgs bosons will be produced via weak boson fusion and the energetic forward jet may suppress the background significantly. However, due to huge $\gamma-g$ scattering background, the potential of LHeC search is still limited. {With a $-80\%$ polarized electron beam of 60~GeV, the signal significance can only reach $2\sigma$ for $ \kappa_{c}\simeq 1.18$ with a 3 ab$^{-1}$ integrated luminosity. In contrast, measurement at the CEPC can reach $5.8\sigma$ for $\kappa_{c}\simeq 1$ with a 2~ab$^{-1}$ of data.}
\end{abstract}

\maketitle
\section{\label{intro}Introduction}	

The observation of Higgs boson decays in di-photon, four-lepton, di-lepton channels \cite{Aad:2012tfa,Chatrchyan:2012xdj}  has provided convincing evidence that the Higgs boson is responsible for generation of the weak gauge boson masses through electroweak gauge symmetry breaking (EWSB). Generation of all charged fermion masses is accommodated in the standard model (SM) as a result of $[U(3)_{Q_{L}}\times U(3)_{u_{R}}\times U(3)_{d_{R}}\times U(3)_{\ell_{L}}\times U(3)_{e_{R}}]$  ($[U(3)]^{5}$) chiral symmetry breaking  as well as EWSB, while the electrically neutral neutrino may contain a Majorana component.   
However, by far, direct measurements of Yukawa couplings have been confirmed at the LHC for only the 3$^{\rm rd}$ generation charged fermions, i.e., 
$h \to b\bar{b}$ decay mode in the associated production (VH) \cite{Sirunyan:2018hoz,Aaboud:2018zhk} and $h \to \tau^{+}\tau^{-}$ with a branching fraction at the percent level measured by ATLAS and CMS individually \cite{Aaboud:2018pen,Sirunyan:2017khh}; 
and the direct measurement of  (t$\bar{\text{t}}$H) \cite{Aaboud:2018urx,Sirunyan:2018hoz}.~With the LHC upgraded to its high-luminosity mode (HL-LHC),  measurements of the Yukawa couplings of the 3$^{\rm rd}$ generation charged fermions are expected to reach an ${\cal O}$(10--20)\% accuracy \cite{atlas2014projections}.

The fermion mass $m_{f}$ can be written as 
\beq
m_{f}= y_{f} v_{0} + \Delta m_{f} \nonumber
\eeq
where $v_{0}$ is the vacuum expectation value of the 125~GeV Higgs boson and $y_{f}$ is the Yukawa coupling from the term $ y_{f} \bar{\psi}_{L}\psi_{R} h$.~$\Delta m_{f}$ is identified as the contribution to $m_{f}$ from physics beyond the standard model (BSM).  
The precision measurement of  $y_{f}$ will then provide a probe of the new physics associated with $\Delta m_{f}$.~Because of the $SU(2)_{L}\times U(1)_{Y}$ gauge structure, the SM charged fermion masses arise from couplings between $SU(2)$ doublets.~In many models such as two-Higgs-doublet models (2HDM)~\cite{Branco:2011iw}, 
$\Delta m_{f}$ is proportional to the corresponding Yukawa coupling $y_{f}$ and  $y_{f}=0$ corresponds to the chiral symmetry limit.~In this case, precision measurement of $y_{b}$ is sufficient to probe the new physics scale. 
However, the coupling $y_{f}$ may not be the only source of chiral symmetry breaking. 
There is evidence for the Yukawa interaction of the 3$^{rd}$ generation charged fermions from direct measurements. The couplings have the form 
\beq
-y_{t}\bar{u}^{3}_{R}Q^{3}_{L} \tilde{h} -y_{b} \bar{d}^{3}_{R} Q^{3}_{L} {h} - y_{\tau}\bar{e}^{3}_{R} \ell^{3}_{L} {h}+h.c. \nonumber
\eeq
The above terms explicitly break the global  $[U(3)]^{5}$ chiral symmetries  of the SM fermions kinetic terms $\bar{\psi}_{L}\cancel{D}_{\mu}\psi_{L}+\bar{\psi}_{R}\cancel{D}_{\mu}\psi_{R}$ down to 
$[U(2)]^{5}$. The most straightforward question is whether the interactions of the first two generation charged fermions with the Higgs are the same as the third generation case. This question is related to the unsolved puzzle of flavor's origin \footnote{ The Applequist-Chanowitz Unitarity $q\bar{q} \to V_{L}V_{L}$  \cite{Appelquist:1987cf,Chivukula:2007gse,Perez:2015aoa}  provided the constraints over all SM fermion mass generation assuming no Yukawa coupling
\bea
\sqrt{s} \lesssim \frac{8\pi v^{2}}{\sqrt{6}m_{c,s,d,u}} \approx 1 \times 10^{3},1 \times 10^{4},2 \times 10^{5},5 \times 10^{5}~\text{TeV}. \nonumber
\eea
The stronger unitarity bound comes from $q\bar{q} \to nV_{L}$ process \cite{Maltoni:2000iq,Dicus:2004rg}, which gives $\sqrt{s} \lesssim 31,52,77,84~\text{TeV}$.}	
. If, for instance, there exists terms
\beq
\bar{Q}^{2}_{L}\gamma^{\mu} Q^{3}_{L} Z^{\prime}_{\mu}+\bar{u}^{2}_{R}\gamma^{\mu} u^{3}_{R} Z^{\prime}_{\mu}, \nonumber
\eeq
then the interaction with $Z^{\prime}$ can mediate the broken $U(1)_{Q^{3}_{L}}\times U(1)_{u^{3}_{R}}\times U(1)_{d^{3}_{R}}$ to the remnant $[U(2)]^{5}$ through loop corrections. Such vector-like flavor violating gauge interaction can be easily realized from a remnant of gauged $SU(3)_{H}$ horizontal symmetry after the Majorana neutrino mass matrix breaking \cite{Wilczek:1978xi,Yanagida:1979as}. 
Direct observation of the Yukawa couplings of the first two generation charged fermions is critically important for solving the above puzzle in the Higgs sector.

 The channel $h \to \mu^{+}\mu^{-}$ as the cleanest decay mode at the LHC is possible to be observed in spite of the 0.2\textperthousand~branching ratio \cite{Han:2002gp,Plehn:2001qg}. ATLAS and CMS have presented that their observed upper limits are 2.9 and 2.2 times the SM prediction, but with very low standard deviation \cite{Aaboud:2017ojs,Sirunyan:2018hbu}. The measurement of $h \to e^{+}e^{-}$ still faces huge challenges due to the tiny coupling, and only gives loose bounds \cite{Altmannshofer:2015qra,Khachatryan:2014aep}. In contrast to $h \to \mu^{+}\mu^{-}$, the first two generation hadronic decay modes of the Higgs not only suffer from low branching ratio, but are also very difficult to be distinguished from QCD backgrounds at the LHC. However, some new methods have been proposed to constrain the Yukawa couplings of the light quarks ($u,d,s$) \cite{Goertz:2014qia,Kagan:2014ila,Koenig:2015pha,Soreq:2016rae,Bishara:2016jga,Yu:2016rvv,Carpenter:2016mwd,Gao:2016jcm,Cohen:2017rsk,Brod:2018lbf}. For $h \to c\bar{c}$ decay mode,  ATLAS has recently presented preliminary direct search \cite{Aaboud:2018fhh} using JetFitterCharm algorithm \cite{atlas2015performance}. A future optimistic bound suggests that 6 times the SM rate at 95\% C.L. is achievable at the HL-LHC \cite{atlas2018prospects}. In principle, there are two practicable methods for probing the Higgs-charm Yukawa coupling ($y_{c}$). One approach is to exploit charm tagging and directly probe the charm Yukawa coupling through $h \to c\bar{c}$ inclusive decay mode. The cleanest channel is the associated production of the Higgs with a vector boson ($VH$) \cite{Delaunay:2013pja,Perez:2015aoa}. Recently, other channels are also proposed for probing the charm Yukawa coupling, such as the gluon-gluon fusion production ($ggH$) $gg \to h \to c\bar{c}\gamma$ \cite{Han:2018juw} and the associated production process $gc \to ch$ \cite{Brivio:2015fxa}, which give $|\kappa_{c}|<8.3$ and $|\kappa_{c}|<3.9$ (95\%~C.L.) at the LHC with $3~\text{ab}^{-1}$, where $\kappa_{c}={y_{c}}/{y^{SM}_{c}}$. The second approach is to  measure rare exclusive decays of the form $h \to MV$, where $M$ denotes a vector meson and $V$ is one of the gauge bosons $W$, $Z$ and $\gamma$. This method is viable for any first or second generation quark \cite{Kagan:2014ila}. In order to extract the charm Yukawa, $h \to J/\Psi\gamma \to \mu^{+}\mu^{-}\gamma$ channel is used in \cite{Bodwin:2013gca,Perez:2015aoa}.

Given the difficulty at the LHC, we study the measurement of  $y_{c}$ at electron-hadron colliders and  positron-electron colliders, e.g., the Large Hadron Electron Collider (LHeC) at CERN and the Circular Electron-Positron Collider (CEPC). The LHeC is constructed by adding one electron beam of 60-140 GeV to the current LHC with the 7 TeV proton beam and a forward detector. As a deep inealastic scattering (DIS) facility, it can be a higgs factory in which  Higgs bosons are mainly produced via vector-boson fusion (VBF) processes \cite{AbelleiraFernandez:2012cc}. It provides some distinctive features of the signal -- a Higgs decays to a $c\bar{c}$ pair in the central region and a jet is produced in the forward direction. CEPC as a positron-electron collider could produce a large number of Higgs bosons through the $e^{+}e^{-} \to Zh$ process at a center-of-mass energy of $\sqrt{s} \sim 240~\text{GeV}$ \cite{CEPCStudyGroup:2018rmc}, which provides the cleanest channel for probing $h \to c\bar{c}$ without other large QCD backgrounds. The fixed $\sqrt{s}$ is also desirable in the reconstruction of the invariant mass of the final states.

The rest of this paper is organized as follows. In section \ref{LHeC}, we assess the physics potential of probing the Higgs-charm Yukawa at the LHeC through analysis of signal and main backgrounds based on studies of differential distributions and kinematic features in search channels, selection cuts and observable reconstruction, and simulation of the signal-to-background ratio ($S/B$) and significance ($Z$). In section \ref{CEPC}, we move on to the positron and electron collider -- CEPC, and discuss the possibility of probing charm Yukawa in different $Z$ boson decay channels.~Finally, we conclude in section \ref{con}.

\section{\label{LHeC}Phenomenology of the charm yukawa at the LHeC}
\subsection{Signal and backgrounds}
\begin{figure}[H]
\centering
\subfloat[]{\includegraphics[width=0.2\textwidth]{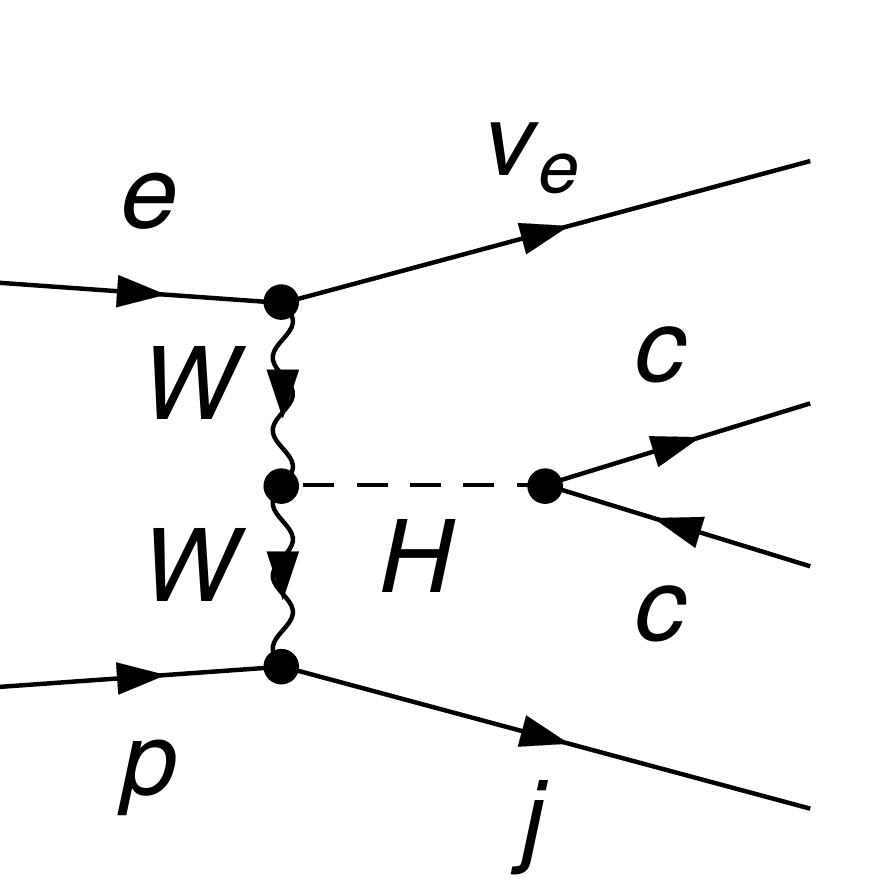}} \quad
\subfloat[]{\includegraphics[width=0.2\textwidth]{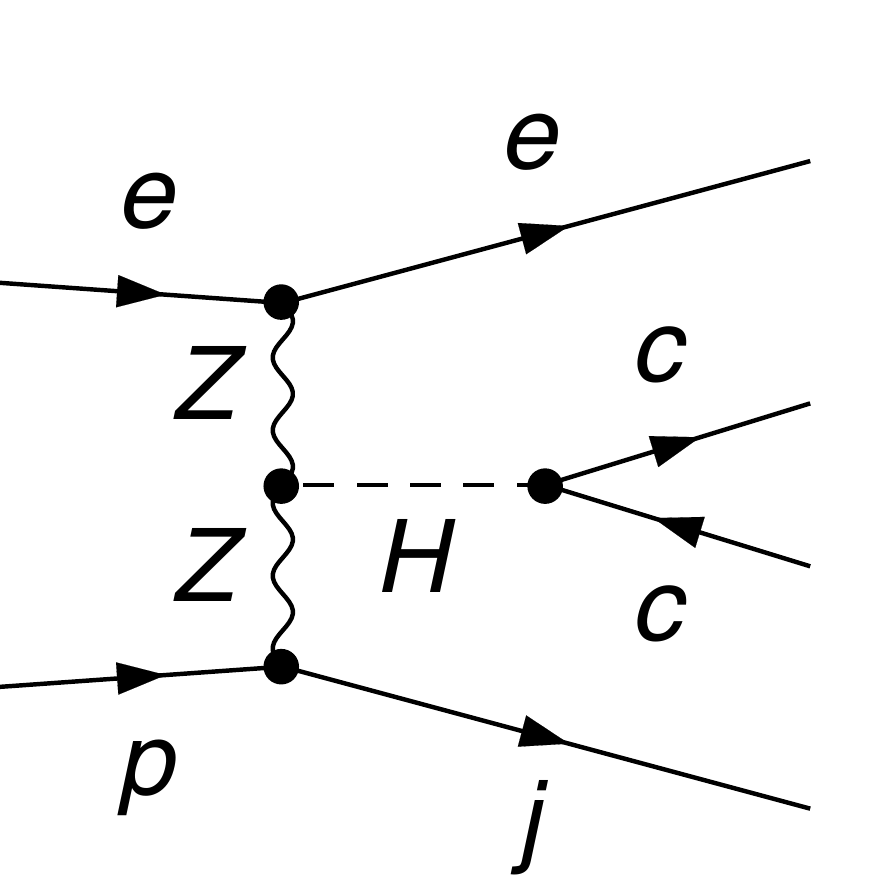}} \\
\subfloat[]{\includegraphics[width=0.2\textwidth]{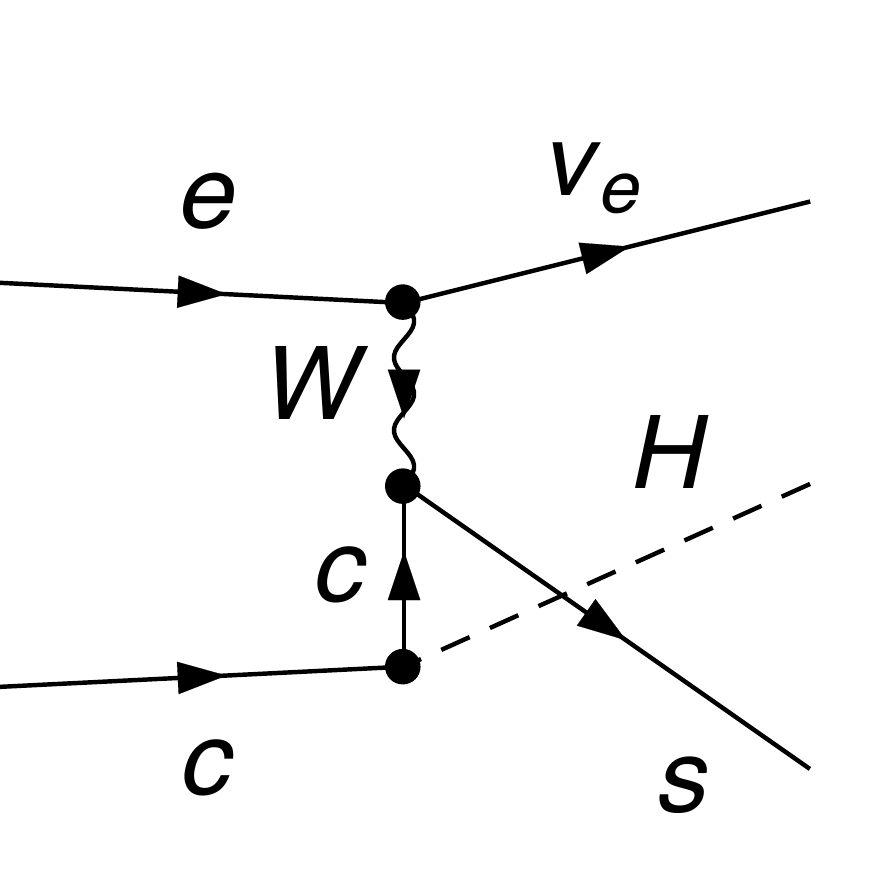}} \quad
\subfloat[]{\includegraphics[width=0.2\textwidth]{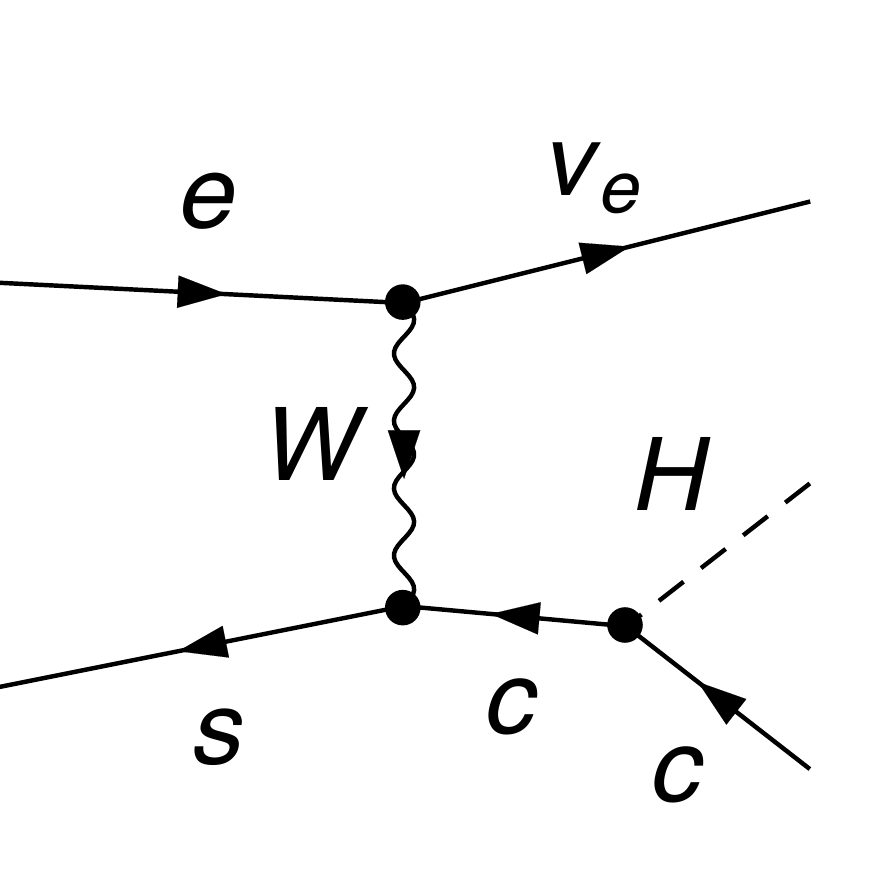}} \quad
\subfloat[]{\includegraphics[width=0.2\textwidth]{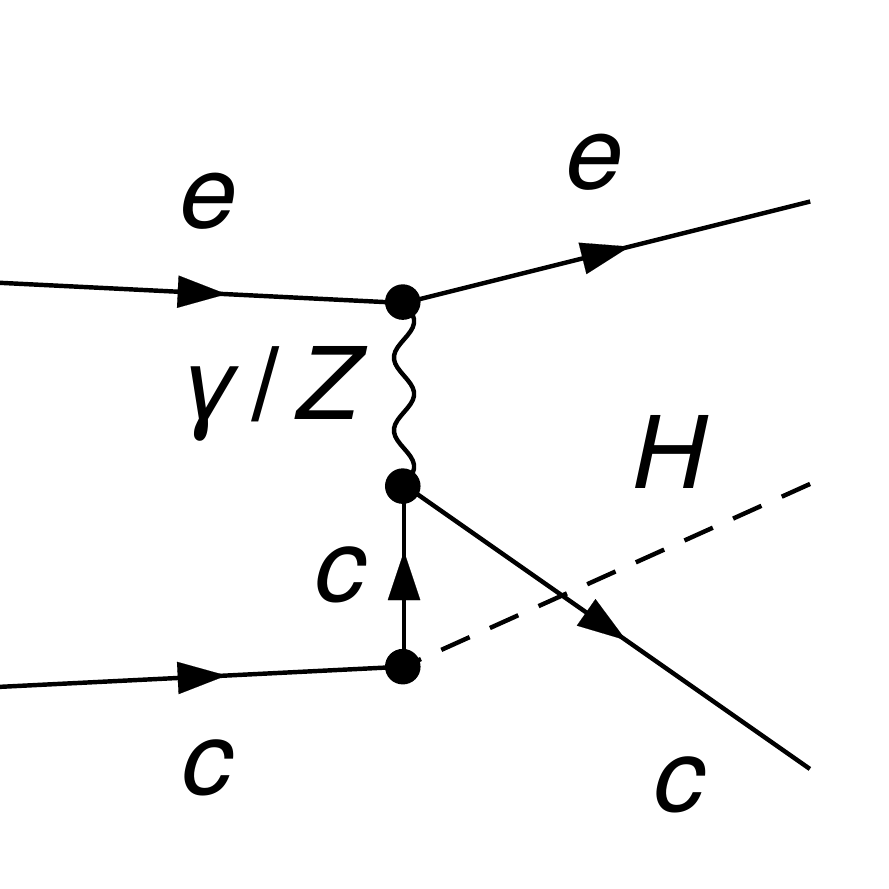}} \quad
\subfloat[]{\includegraphics[width=0.2\textwidth]{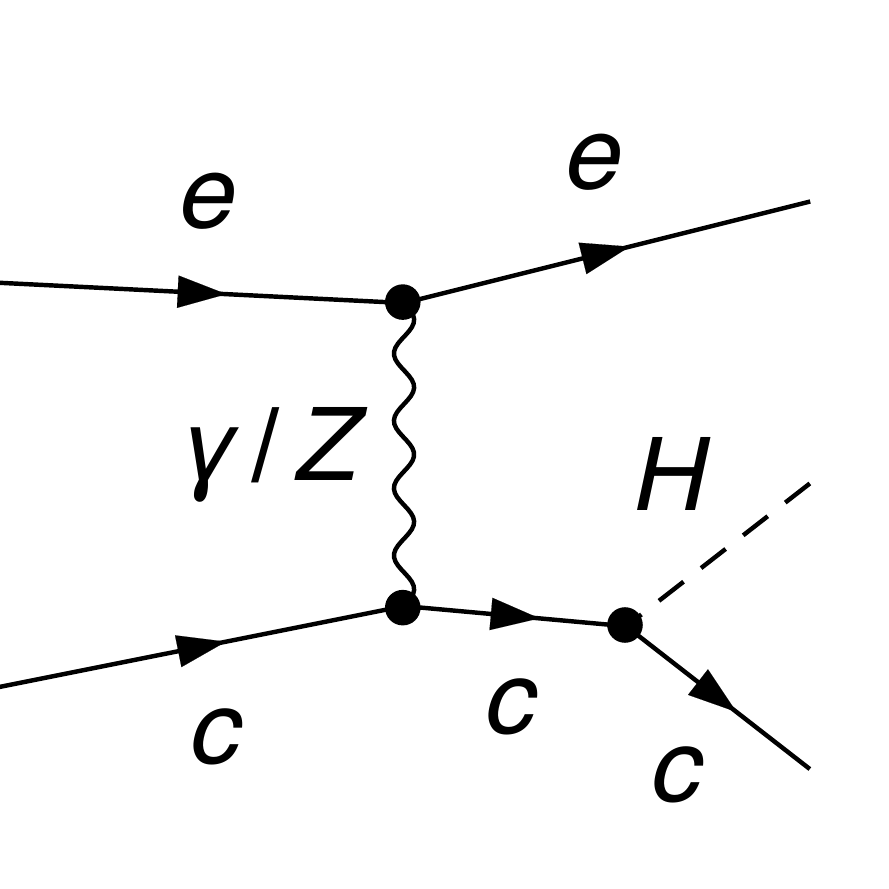}} \\
\subfloat[]{\includegraphics[width=0.2\textwidth]{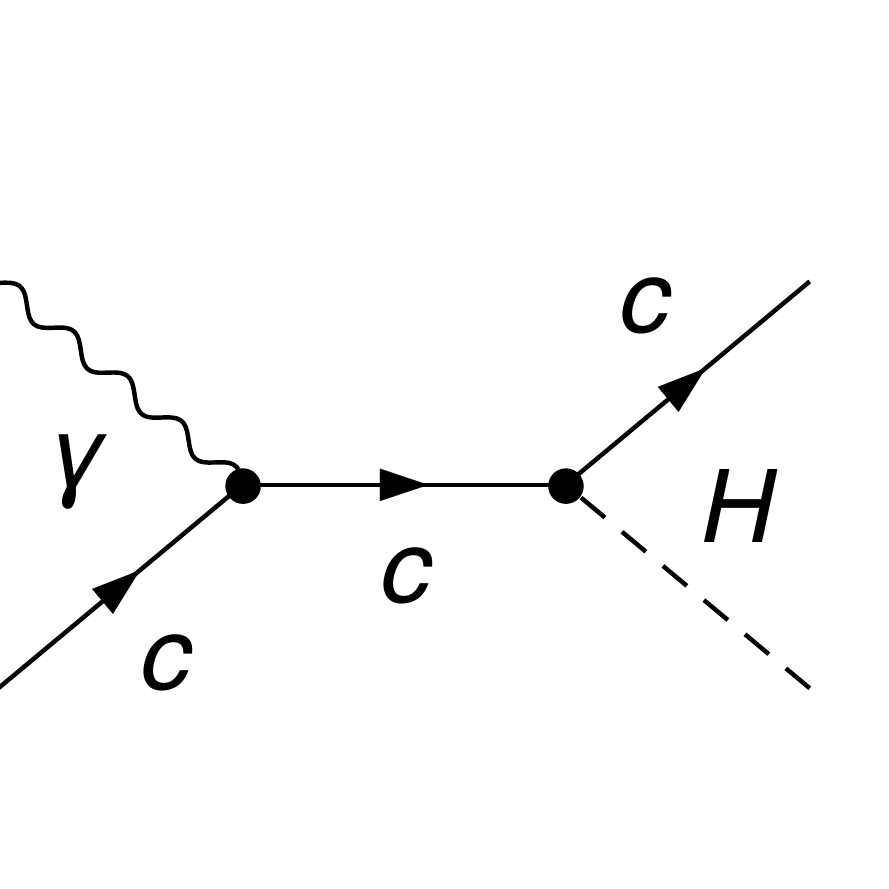}} \quad
\subfloat[]{\includegraphics[width=0.2\textwidth]{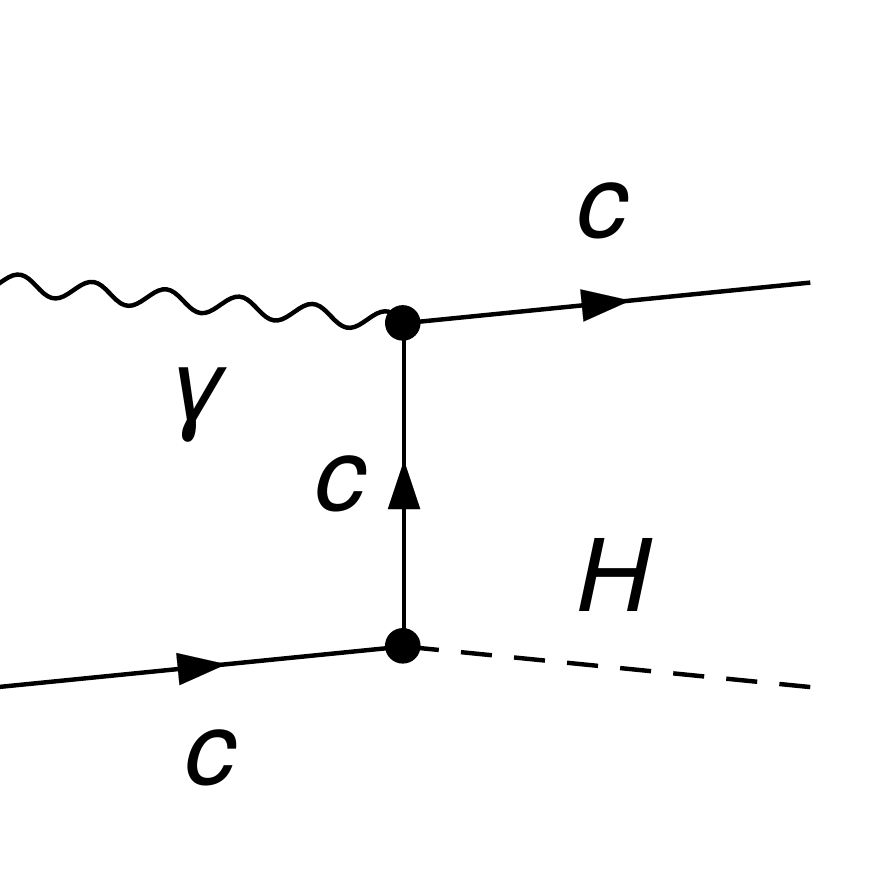}} \quad
\subfloat[]{\includegraphics[width=0.2\textwidth]{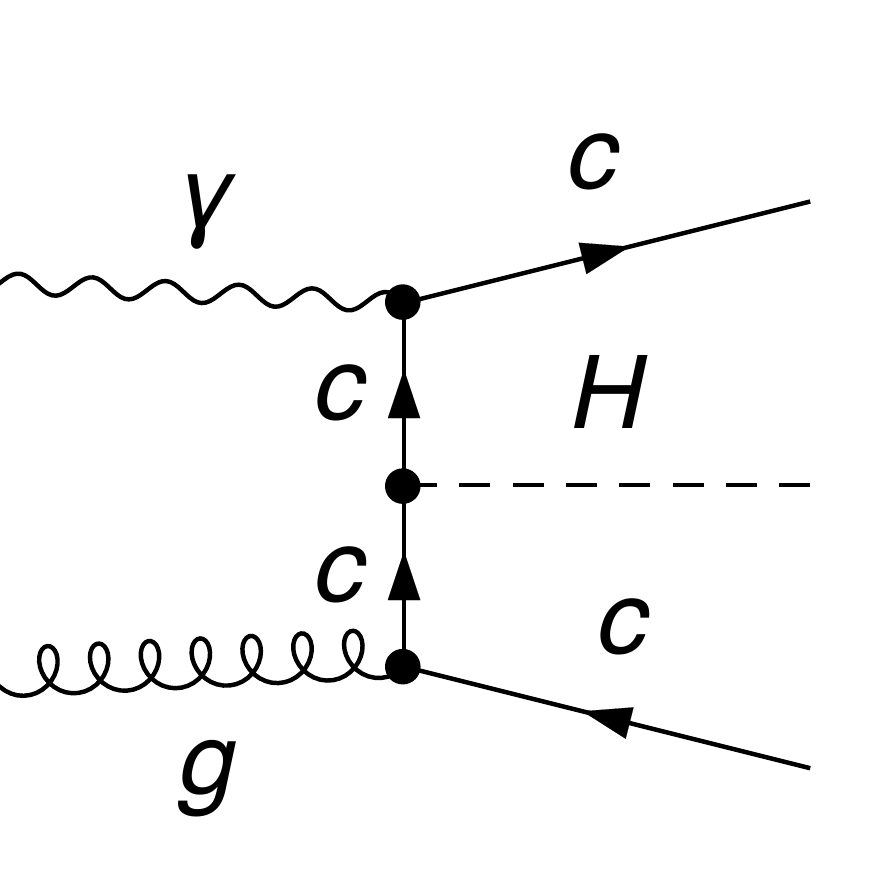}} \quad
\subfloat[]{\includegraphics[width=0.2\textwidth]{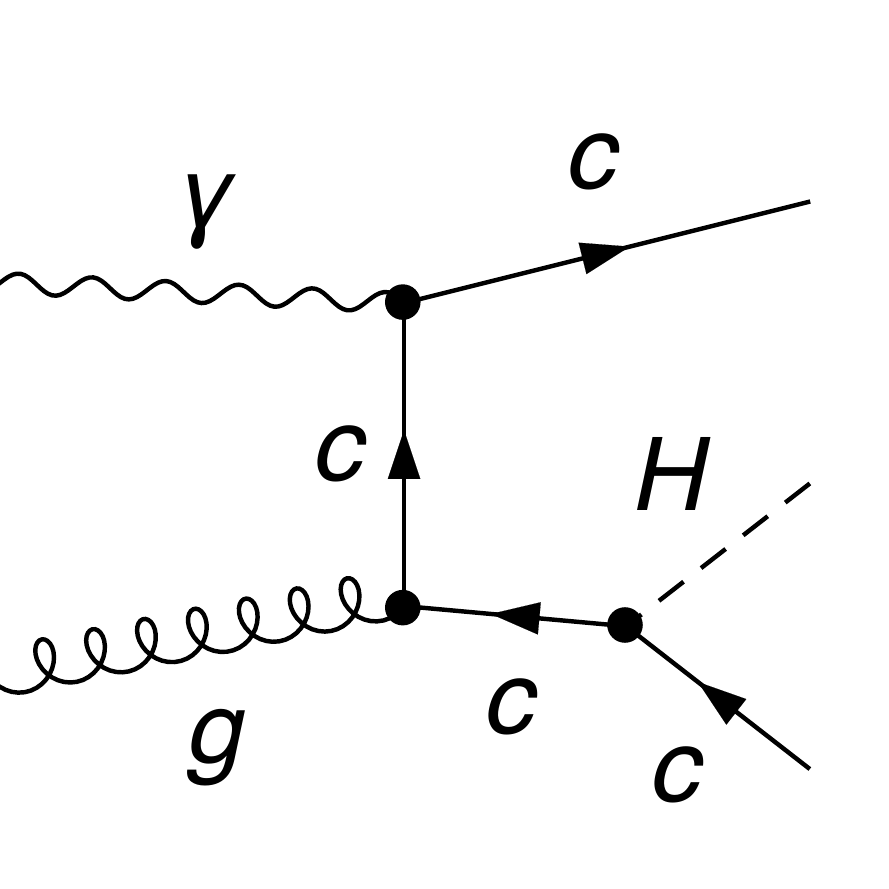}} \\
\subfloat[]{\includegraphics[width=0.2\textwidth]{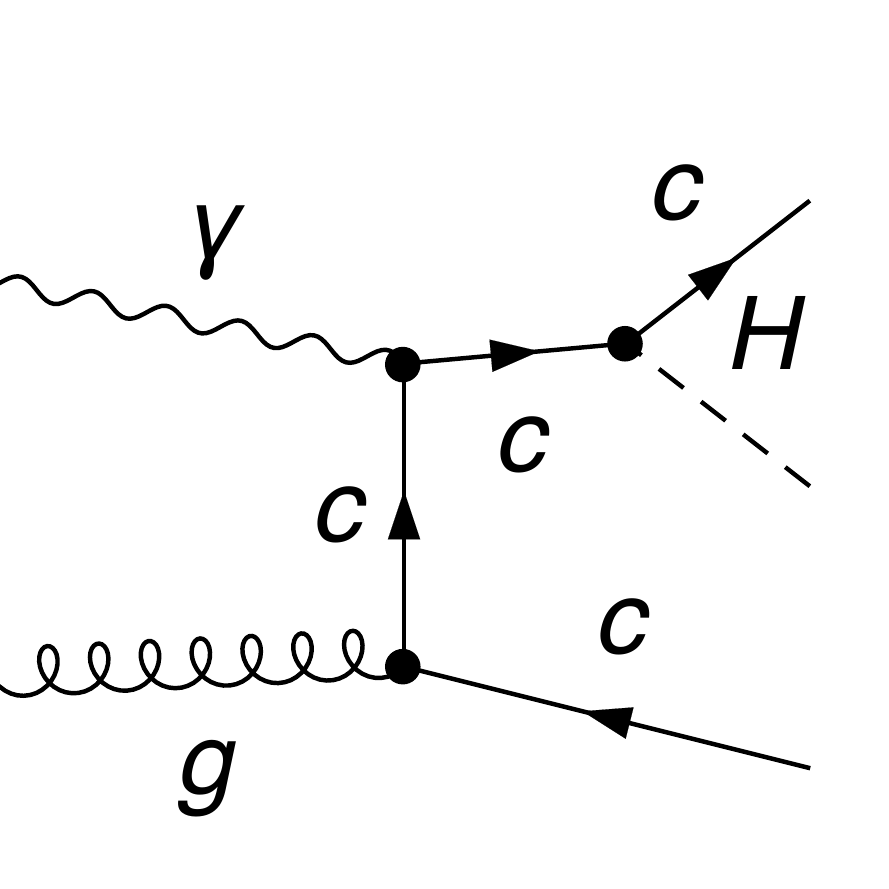}} \quad
\subfloat[]{\includegraphics[width=0.2\textwidth]{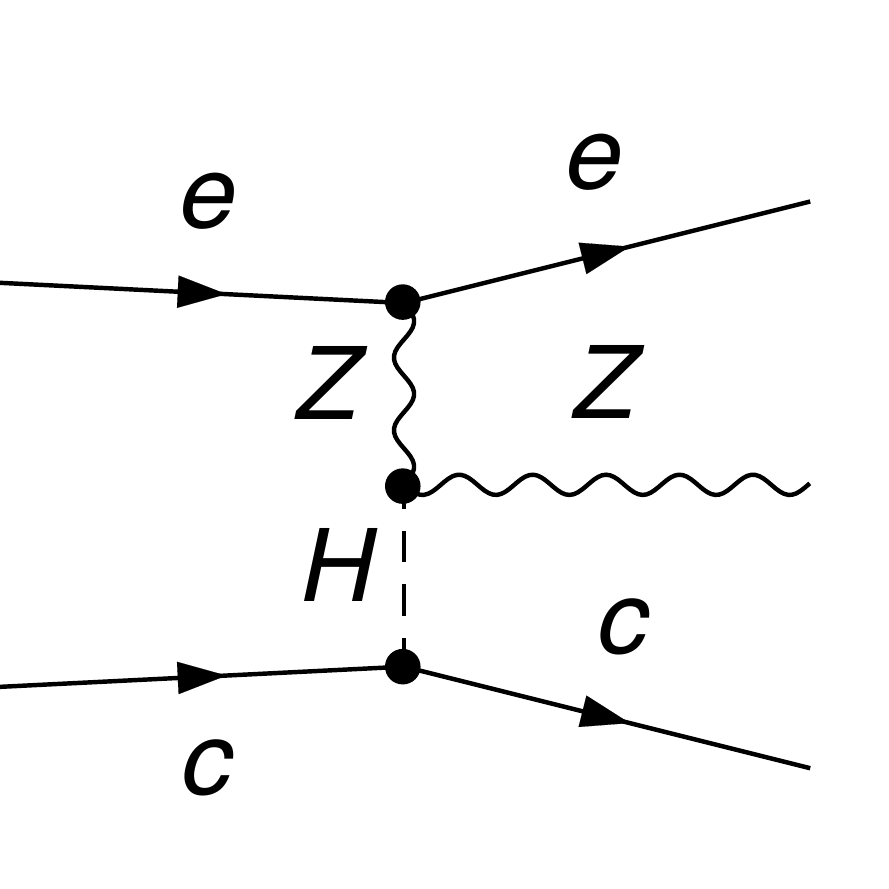}} \quad
\subfloat[]{\includegraphics[width=0.2\textwidth]{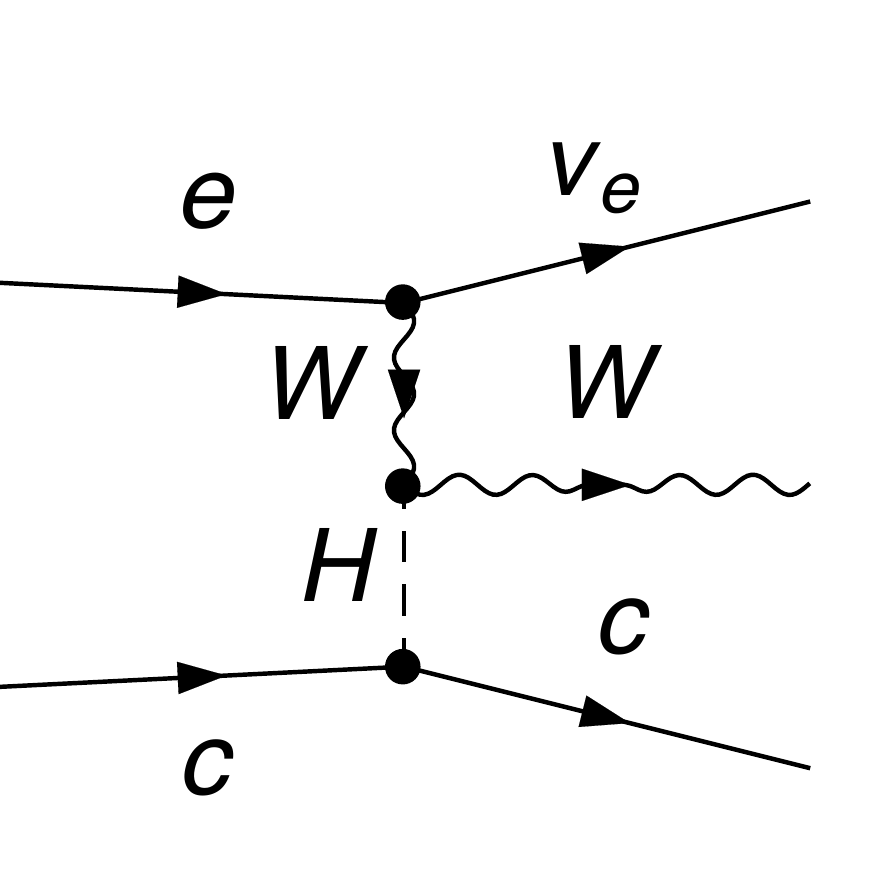}} 
	\caption{Processes containing the Higgs-charm Yukawa coupling at the LHeC. An initial $\gamma$ indicates that it is from the collinear radiation of the incoming electron, for which an effective photon approximation (EPA) \cite{Drees:1994zx} can be used in the calculation. $g$ is the gluon from the proton beam. $H$ is the Higgs boson in the SM.}
\label{process}
\end{figure}

At the $e-p$ collider, the Higgs-charm Yukawa coupling could be probed through three different modes shown in Fig.\ref{process}. (a) and (b) are the vector-boson fusion (VBF) processes used to measure the charm Yukawa coupling via $h \to c\bar{c}$ decay directly. (c)--(k) are the direct production of Higgs bosons from charm quarks, with cross sections about ${\cal O}$($10^{-2}$--$10^{-4}$)~{fb} and can not be measured precisely at the LHeC. (l) and (m) are processes with Higgs boson exchanged between a charm quark and a gauge boson. We could study the interference between these processes and other $e^{-}W+c/\nu_{e}Z+c$ final states to probe the Higgs-charm Yukawa coupling. But the tiny cross sections of ${\cal O}$($10^{-4}$)~{fb} is a big challenge. Therefore, as stated in the introduction, we focus on the dominant Higgs boson production mode, vector-boson fusion, followed by charm quark inclusive decay
\bea
e^{-} p \to \nu_{e} h j , h \to c\bar{c}.  \nonumber
\eea
We drop the neutral current process for its smaller cross section and additional backgrounds, and require the forward jet to be a light-jet ($u,s,d$ and $g$)\footnote{This constraint would reduce the cross section by less than 1\%. We expect that the effect is insignificant.}. {Since only left-handed electrons from the initial beam contribute in this process, we tentatively set the electron beam with polarization of -80\% to improve the signal production rate. }~By the way $j$ only denotes a light-jet unless otherwise noted. c-jet and b-jet are indicated as $c$ and $b$ explicitly.

The Monte-Carlo events are generated by {\it MadGraph5\_v2.6.4}~\cite{Alwall:2014hca} at parton level.~We use {\it Pythia6.420} \cite{Sjostrand:2006za}\footnote{{The $k_T$ clustering jet algorithm is implemented by {\it Pythia6.420}.}} and {\it Delphes3.3.0} \cite{deFavereau:2013fsa}\footnote{{Because the LHeC detector card is not finalized, we simply use the published CEPC card for {\it Delphes3.3.0} to simulate the detector. There is a preliminary version of the LHeC detector card.~We compared the observables reconstructed using this test card and the CEPC card, and found that the difference is insignificant.}} for parton shower and detector simulations respectively. At parton level, we impose the following basic cuts 
\begin{itemize}
\item $p_{T}^{j}>20~\rm{GeV}$,
\item $p_{T}^{\ell}>5~\rm{GeV}$,
\item $|\eta_{\ell,j}|<5$,
\item $\Delta{R}_{j \ell}>0.4$ and $\Delta{R}_{\ell \ell}>0.4$~,
\end{itemize}
where $j$ denotes all kinds of hadronic jet ($j=$ light-jet, c-jet and b-jet), $\ell$ is the lepton ($e, \mu$). It has been argued that the $p^{\ell}_{T} >5$~GeV requirement would not break the convergence of the perturbative calculation \cite{Degrande:2016aje,Han:2018rkz}. When $E_{e}=60~\text{GeV}$ and $E_{p}=7 ~\text{TeV}$, the total cross section of WBF production can be as large as 84.83 fb. Quoting the updated calculation of the branching ratio of the Higgs boson hadronic decay in the SM \cite{pdg2018}, $Br^{SM}(h \to c\bar{c}) \approx 2.9\%$. 
The $\sigma(e^{-} p \to \nu_{e} h j , h \to c\bar{c})=84.83~\text{fb} \times Br^{SM}(h \to c\bar{c}) =2.47$ fb is within the detector resolution at the LHeC.

The analysis of the background is performed as in \cite{Han:2009pe}. The leading irreducible background comes from $\slashed{E}_{T}+\text{muliti-jets}$. In order to distinguish c-jet from others in the final states, we classify all jets into three categories: light-jet, c-jet, and b-jet in the background, e.g. $\nu_{e}jjj,\nu_{e}ccj$ and so on. Hence the irreducible backgrounds are  
\bea
e^{-} p \to \nu_{e} j j j,~\nu_{e} c \bar{c} j,~\nu_{e}b \bar{b} j,~\text{etc}.\nonumber 
\eea
These backgrounds include some subprocesses from on-shell particles decay $h \to b\bar{b},~W \to jj,cj$, fusion subprocesses with gauge bosons exchanged, and the interference between them.

Among the reducible backgrounds, the largest would be the photo-production processes, which can be classified into two categories:
\bea
&&\gamma p \to j j j,~c c j,~b b j,~\text{etc}., \nonumber \\
&&\gamma p \to t \bar{t}, \nonumber
\eea
where $\gamma$ is from the EPA and $p$ denotes the initial parton. The first category includes the photon-quark and photon-gluon fusion, while the second  only has the photon-gluon fusion subprocess because of the production of the on-shell top quark in the final states. All decay modes of the top quark are included when generating this process.

The second largest reducible background is the single top quark production via WBF process followed by top quark decay:
\bea
e^{-} p \to \nu_{e} b \bar{t} j.  \nonumber
\eea
Here we restrict the forward jets to light-jets just like in the signal case, since the c-jet or b-jet appears in the final state only when the initial parton from the proton is charm or bottom, which is insignificant.~{A summary of the unpolarized and polarized cross sections of the signal and backgrounds is given in Table.\ref{xsection}. It is shown that the backgrounds are far larger than the signal after the -80\% polarized electron beam has been set}. Effective cuts are needed to reduce the huge backgrounds. 
{Since the signal we focus on is a charged current process and only left-hand electrons contribute, the sole effect of the polarized electron beam is to increase the
electron luminosity by 80\% compared to the unpolarized case.~The same increase also occurs to the $\slashed{E}_{T}+\text{multi-jets}$ and $ \nu_{e} b \bar{t} j $ backgrounds, as is reflected by the differences in the polarized and unpolarized cross sections. In contrast, the $\gamma p$ backgrounds are not sensitive to the polarization of the electrons. In the following we will explore these processes when the initial unpolarized electron beam has the energy of 60 GeV and generalize this analysis to the -80\% polarization case.} 
\begin{table}[H]
\renewcommand\arraystretch{1.0}
\centering
\begin{tabular}{|c||c|c|c|c|c|}
\hline
\hline
S\&B & signal & $\slashed{E}_{T}+\text{multi-jets}$ & $\text{multi-jets}(\gamma p)$ & $t\bar{t}(\gamma p)$ & $ \nu_{e} b \bar{t} j $ \\
\hline
unpolarization & 2.47 & 8073.45 & 228485.6 & 2385.2 & 122.54 \\
\hline
-80\% polarization & 4.40 & 14532.2 & 228423.6& 2383.6& 220.57 \\
\hline
\hline
\end{tabular}
\caption{Cross-sections (in fb) for the signal and four main backgrounds when $E_{e}=60~\text{GeV}$ with/without polarization of -80\%.}
\label{xsection}
\end{table}

\subsection{Selection cuts}
\begin{figure}[H]
\centering
\subfloat[The invariant mass $M(c,c)$ of two c-jets]{\includegraphics[width=0.4\textwidth]{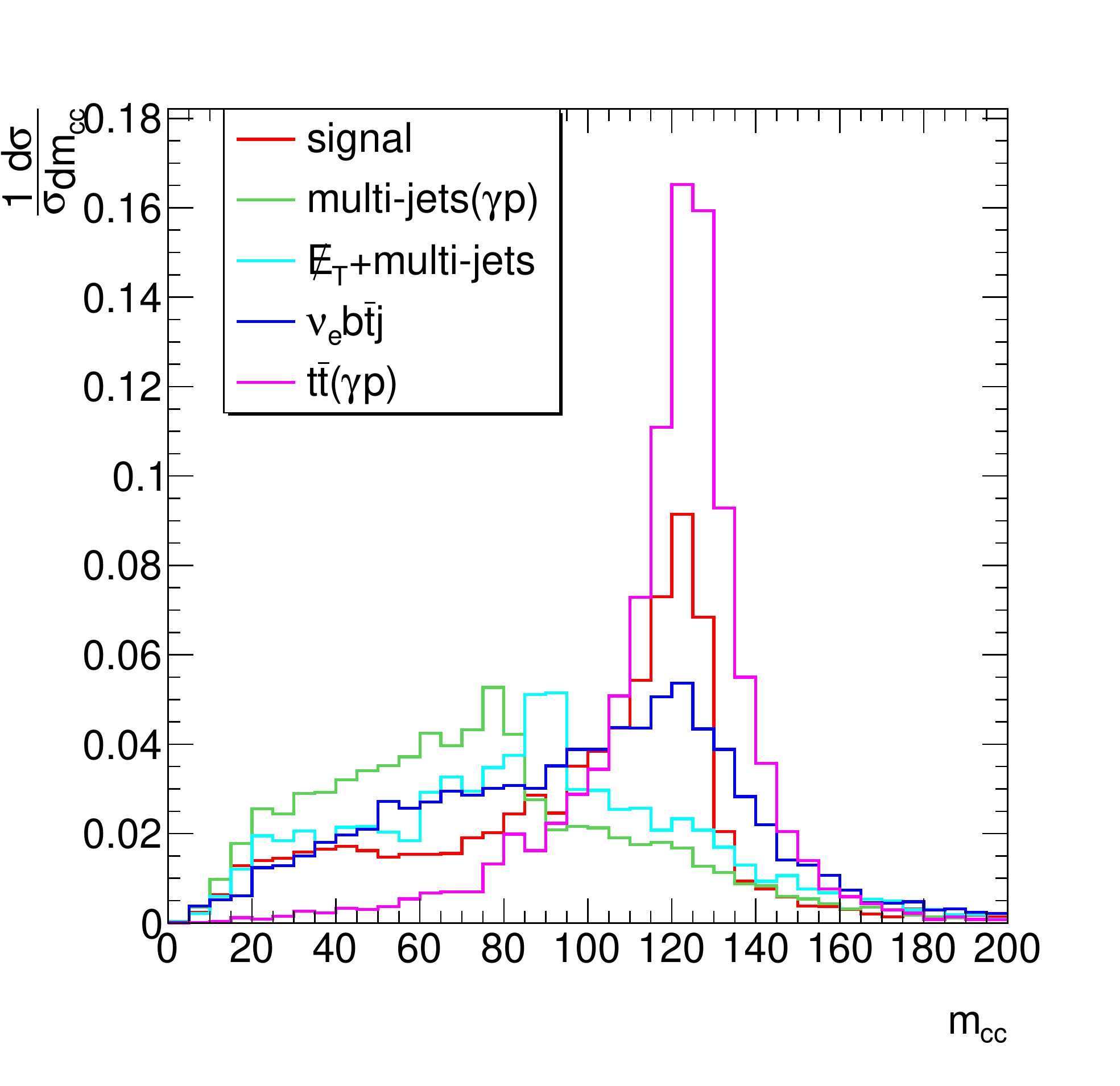}}
\subfloat[The invariant mass $M(c,c,j)$ of the two c-jets and the forward light-jet]{\includegraphics[width=0.4\textwidth]{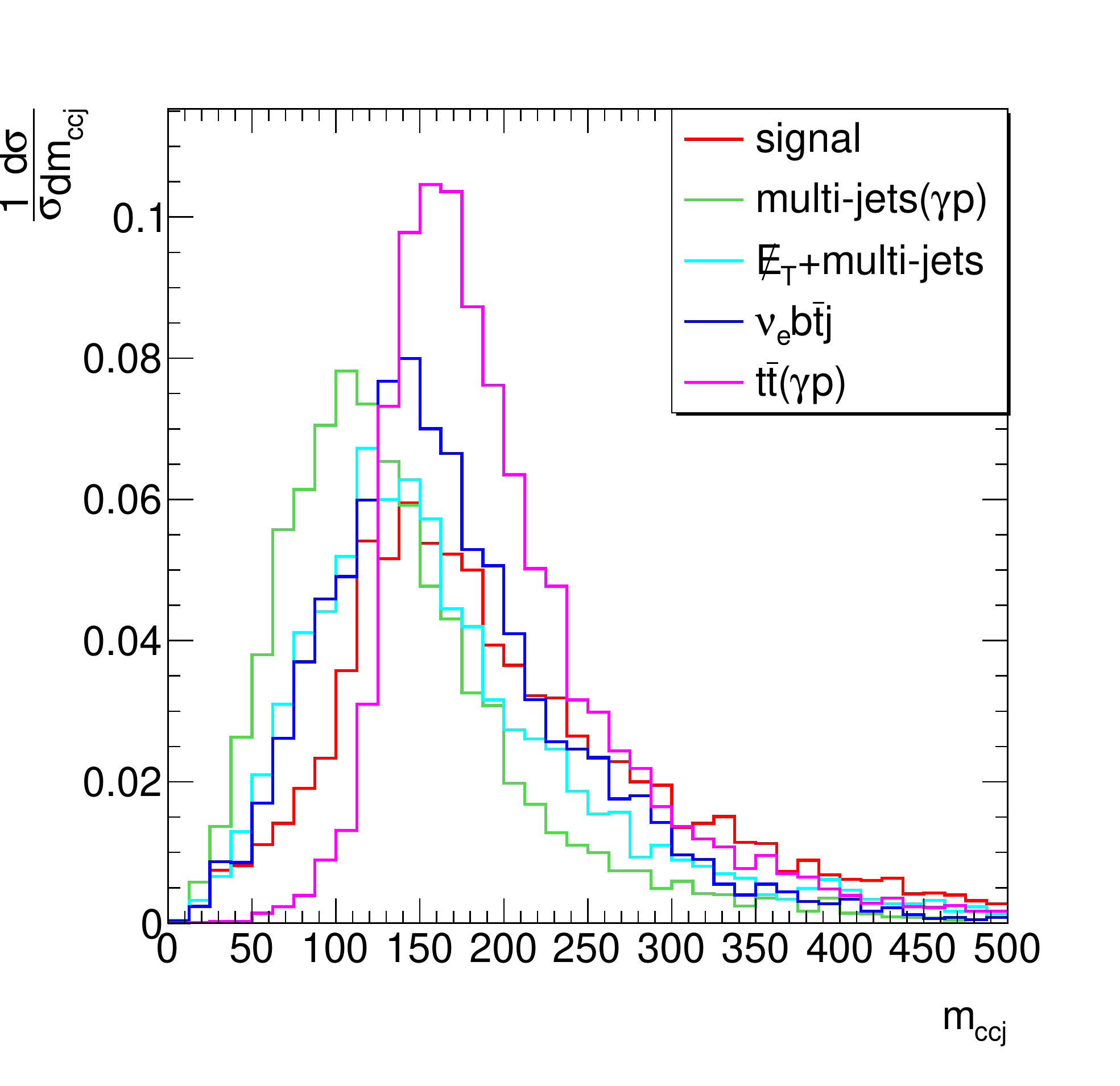}} 
\caption{The normalized distribution of the invariant mass $M(c,c)$ (left panel) and $M(c,c,j)$ (right panel) after full detector simulation.}
\label{masslhec}
\end{figure}

Restricting the forward jet to be light for the signal induces a deviation in comparison with $\slashed{E}_{T}+\text{multi-jets}$ and photo-production processes.~Then a {central jet veto can reduce the backgrounds like the photo-production which contain extra QCD radiations in the central region.}~The obvious feature of the signal is that the distribution of the invariant mass $M(c,c)$ of the two c-jets from Higgs boson decays has a peak at 125 GeV. In order to extract the right jets to construct the invariant mass $M(c,c)$, we can choose the two of the remaining jets, whose $M(c,c)$ is closest to  the 125 GeV peak, after excluding the one with the maximal pseudorapidity. Of course the distribution of the $\slashed{E}_{T}$ is different between the signal and the photo-production backgrounds because the $\gamma p \to \text{multi-jets}$ process lacks missing transverse energy at parton level, and the $\gamma p \to t\bar{t}$ only produces neutrinos from top quark semi-leptonic decays that have relatively small branching ratios. Meanwhile, a veto on events with extra leptons could suppress the backgrounds containing top quarks. {In Fig.\ref{masslhec}, distributions of the $M(c,c)$ and $M(c,c,j)$ are plotted respectively, where the red line corresponds to the signal, the green, cyan, blue and magenta lines correspond to the four main backgrounds.~We found the $t\bar{t}(\gamma p)$ background has the largest overlap with the signal in these distributions.~The signal peak of the $M(c,c)$ distribution has a small deviation from the Higgs mass pole from the full detector simulation.}~Therefore, we adopt the following selection cuts criteria at the LHeC:
\begin{itemize}
\item[i.] We require the jet with the maximal pseudo-rapidity is the light-jet (forward light-jet) and $|\eta_{j}|>2.6$.
\item[ii.] A veto on events with any soft jets in the central region with $p^{j}_{T}<10~\rm{GeV}$ and $|\eta_{j}|<2.4$.
\item[iii.] The invariant mass of the remainning two c-jets in the final state $110~\rm{GeV}<$$M(c,c)$$<135~\rm{GeV}$.  
\item[iv.] A veto on events with extra leptons with $p^{\ell}_{T}>5~\rm{GeV}$. 
\item[v.] The missing transverse energy cut:~$\slashed{E}_{T}>20~\rm{GeV}$ . 
\item[vi.] The transverse momentum of the leading c-jet ~$p^{c_{1}}_{T}>70~\rm{GeV}$ .  
\item[vii.] The invariant mass of the two c-jets and the forward light-jet in the final state $M(c,c,j) >250~\rm{GeV}$. 
\end{itemize}

So far we only set the kinematic cuts and have not considered the tagging and mis-tag efficiency.~In fact, tagging system is crucial for the search of $h\to c \bar{c}$. A high $c$-tagging efficiency and a low misidentification rate could reduce the multi-jets background dramatically. For instance, the $\gamma p \to j j j$ process contributes 55\% to the total cross section of the $\text{multi-jets}(\gamma p)$ background, whereas the rate of the light-jet faking to c-jets in the final state is only at percent level.~Even though some background events pass all kinematic selection cuts, they are still suppressed by a mis-tag efficiency of $\mathcal{O}(10^{-1}-10^{-2})$ at least. Based on the ATLAS and previous measurements of the charm Yukawa \cite{Aaboud:2018fhh}, we assume the $c$-tagging {efficiencies} and b-jet, light-jet mis-tag {rates} as in Table.\ref{tag} at different operating points. The cut-flow for the signal and background events is presented in Table.\ref{cut-flow}.
\begin{table}[H]
\renewcommand\arraystretch{1.0}
\centering
\begin{tabular}{|c||c|c|c|}
\hline
\hline
{Efficiency} & $\epsilon_{c}$ & $\epsilon_{b\to c}$ & $\epsilon_{j\to c}$  \\
\hline
 i &20\% & 5\% & 3.3\% \\
 ii &30\% & 11\% & 3.3\% \\
 iii &41\% & 25\% & 5\% \\
\hline
\hline
\end{tabular}
	\caption{$\epsilon_{c}$, $\epsilon_{b\to c}$ and $\epsilon_{j\to c}$ denote $c$-tagging {efficiencies}, b-jet mis-tag {rates} and light-jet mis-tag {rates} at different operating points respectively. {iii is the working point used in \cite{Aaboud:2018fhh}}. }
\label{tag}
\end{table} 

\begin{table}[H]
\renewcommand\arraystretch{0.7}
\centering
\begin{tabular}{|c||c|c|c|c|c|c|}
\hline
\hline
	cuts & signal & $\slashed{E}_{T}+\text{multi-jets}$ & $\text{multi-jets}(\gamma p)$ & $t\bar{t}(\gamma p)$ & $ \nu_{e} b \bar{t} j $   \\
\hline
\hline
	basic cuts & 2470 & 8073450 & 228485600 & 2385200 & 122540  \\
\hline
	forward light-jet and $|\eta_{j}|>2.6$ & 1947.35 & 3576320 & 114214000 & 1621940 & 73210  \\
\hline
	central jets veto & 1435.2 & 2578530 & 78578900 & 712032 & 44072.4  \\
\hline
	$110~\rm{GeV}<$$M(c,c)$$<135~\rm{GeV}$ & 439.74 & 303781 & 7506950 & 361326 & 11985.8  \\
\hline
	veto on events with extra leptons & 413.36 & 151891 & 3753480 & 180663 & 5992.9  \\
\hline
	$\slashed{E}_{T}>30~\rm{GeV}$ & 372.02 & 136701 & 187674 & 162597 & 5393.61  \\
\hline
	$p_{T}^{c_{1}}>70~\rm{GeV}$ & 248.77 & 87625.6 & 55703.3 & 131293 & 2958.16  \\
\hline
	$M(c,c,j) >250~\rm{GeV}$ & 141.43 & 14079.3 & 5981.05 & 4916.5 & 992.14  \\
\hline
	tagging i & 5.66 & 70.34 & 151.04 & 103.64 & 20.12 \\
	tagging ii & 12.73 & 154.66 & 333.90 & 200.95 & 46.78 \\
	tagging iii & 23.78 & 570.68 & 668.34 & 296.65 & 138.93 \\

\hline
\hline
\end{tabular}
\caption{Cut-flow of the signal and background events at 60~GeV electron beam energy LHeC with unpolarization when $\mathcal{L}= 1~\text{ab}^{-1}$. The last row presents the number of events after kinematic cuts times the corresponding tagging efficiency at different points. }
\label{cut-flow}
\end{table} 
{After the kinematic cuts, only $\mathcal{O}(10^{-4})$ of the photo-production background, and $\mathcal{O}(10^{-2})$ of the $\slashed{E}_{T}+\text{multi-jets}$ and $ \nu_{e} b \bar{t} j $ backgrounds survive,~while still around 5\% of the signal remains.~{The forward light-jet tagging, central jets veto and $M(c,c)$ reconstruction reduce the photo-production backgrounds by approximately one order after excluding a large rate of QCD jets produced in the central region.~The invariant mass cut of the two c-jets is efficient for all backgrounds as we expected though there is some overlap between the signal and $t\bar{t}(\gamma p)$ background.} The missing transverse energy and $p_{T}^{c_{1}}$ cuts suppress the $\text{multi-jets}(\gamma p)$ background by approximately two orders. At last, after implementing the assumed tagging efficiency in Table.\ref{tag}, the remaining background events have a dramatic decline and become much closer to the signal. }

\subsection{Results}
\label{LHeC_results}
\begin{figure}[H]
\centering
\subfloat[$Z$]{\includegraphics[width=0.45\textwidth]{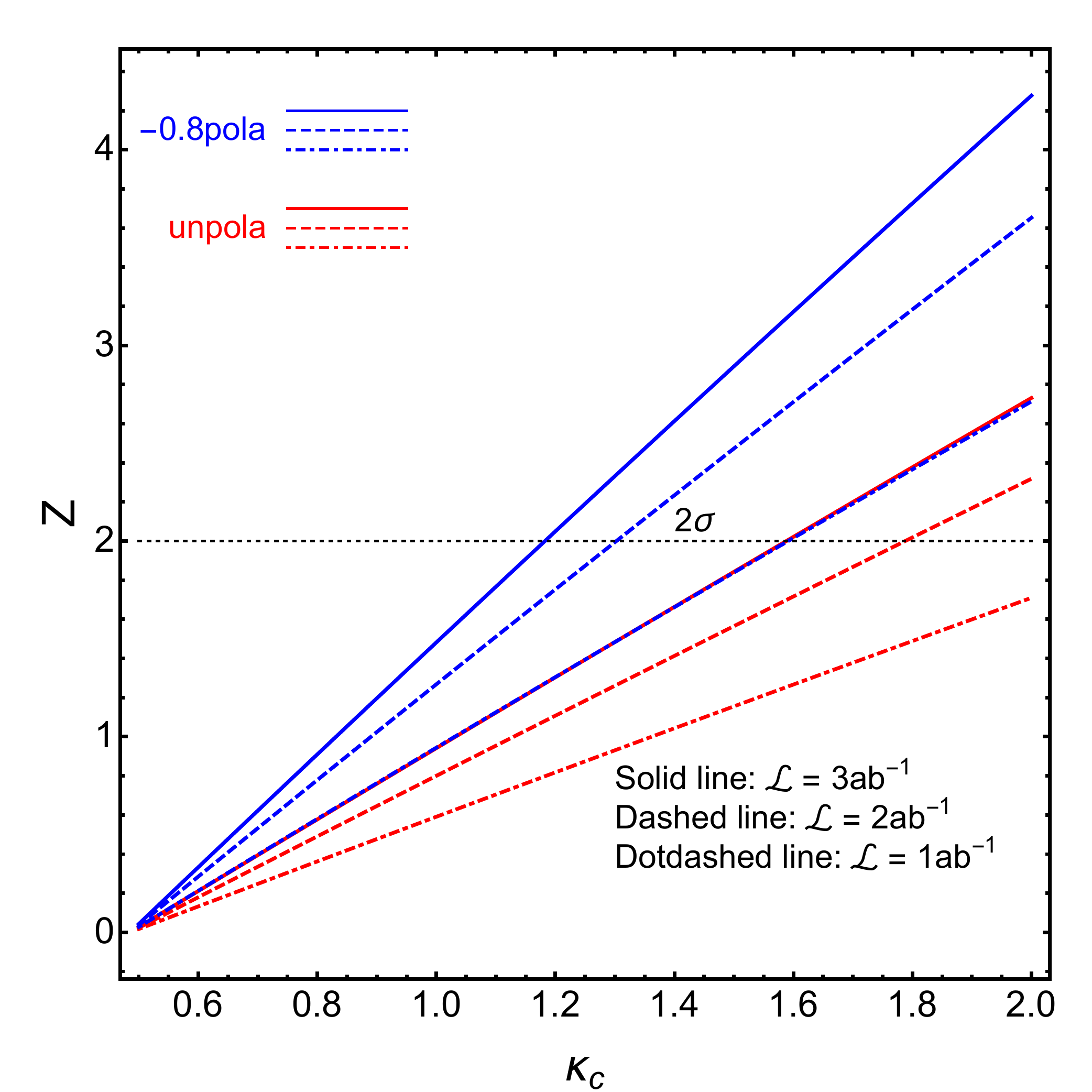}} \quad
\subfloat[$S/B$]{\includegraphics[width=0.45\textwidth]{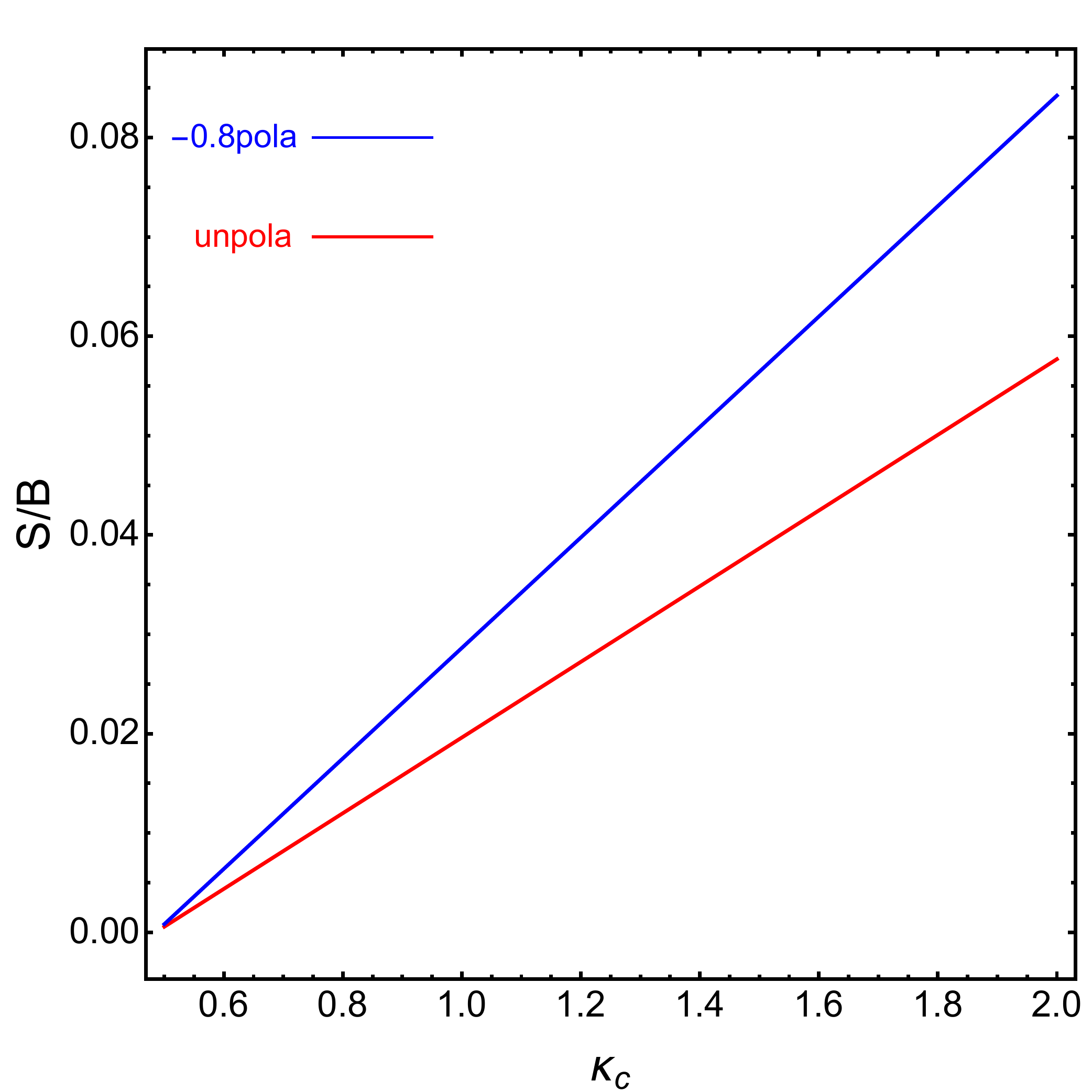}}
\caption{(a): the significance $Z$ varying with $\kappa_{c}$ at the LHeC; (b): the signal-to-background ($S/B$) varying with $\kappa_{c}$ at the LHeC. The energy of the electron beam is 60 GeV. The tagging iii is used.}
\label{significance}
\end{figure}

We calculate the signal significance $Z$ through
\bea
Z=\frac{S}{\sqrt{S+B}} \nonumber
\eea
{where $S$ repsesents the number of signal events. The overall background $B$ including the 1\% systematic error reads $B=\Sigma_{i}B_{i}+\Sigma_{i}(0.01B_{i})^{2}$ ($i = \slashed{E}_{T}+\text{multi-jets}$,~$\text{multi-jets}(\gamma p)$,~$t\bar{t}(\gamma p)$ and $ \nu_{e} b \bar{t} j $). To estimate the effect of varying the charm Yukawa coupling on the $h \to c \bar{c}$ branching ratio, we use the following formula: }
\bea
Br(h \to c\bar{c})\equiv \frac{\Gamma_{h\to c\bar{c}}}{\Gamma_{tot}} \approx Br^{SM}(h \to c\bar{c})\bigg(1+2\delta\kappa_{c}-2Br^{SM}(h \to c\bar{c})\delta\kappa_{c}\bigg) \nonumber
\eea
where ${\Gamma_{h\to c\bar{c}}}$ and ${\Gamma_{tot}}$ are the partial width of the Higgs boson decay to the charm quark and the total width respectively, $Br^{SM}(h \to c\bar{c}) \approx 2.9\%$ is the branching ratio of the Higgs boson decay to the charm quark in the SM, and $\delta\kappa_{c}=\kappa_{c}-1$. The significance ($Z$) and signal-to-background ($S/B$) dependence on $\kappa_{c}$ {are plotted in Fig.\ref{significance} with various polarizations of the initial electron beam and integrated luminosities.} {As an example, the plots are made assuming {the tagging efficiencies} from the last row of Table.\ref{tag}.} {The red and blue lines correspond to the unpolarized and -80\% polarized initial electron beam respectively. It is clearly shown that with the 3 ab$^{-1}$ integrated luminosity and polarized electrons (blue solid line), the significance can reach $2\sigma$ (95\% C.L.) at $\kappa_{c}=1.18$. In contrast, the signal only gets to a $1.3\sigma$ significance at the same $\kappa_{c}$ if the initial electron beam is unpolarized (red line). In the polarized case, the signal-to-background also improves prominently because the electron luminosity increases for the signal while the largest $\gamma p$ backgrounds are essentially unaffected. At $\kappa_{c} \sim 1.38$, $S/B$ reaches to 5\% when the electrons are -80\% polarized, while in the unpolarization case, $\kappa_{c}$ needs go up to 1.8 in order to get the same signal-to-background. Therefore, the initial polarization is helpful for probing the charm Yukawa at LHeC. }

We anticipate further improvements in the cut criteria and tagging {efficiencies} that lead to larger significance and signal-to-background.\footnote{{For instance, it is possible to tag also the final state electron to further suppress the background from the photo production processes. A study in {progress} \cite{Photo} shows that the electron tagging in our case could increase the significances presented above by approximately a factor 1.2.}} {More sophisticated and systematic analysis {(e.g., the machine learning method)} may also be needed for precise measurements of the charm Yukawa coupling \cite{Klein:2018rhq,Abada:2019lih}.} 
Nevertheless, the crude estimation for the sensitivity of the charm Yukawa coupling shows a significant improvement over the results of the LHC \cite{Aaboud:2018fhh}. {Also, our detailed catalog and analysis of the backgrounds for probing charm Yukawa provide useful information for further studies of the subject.}

\section{\label{CEPC}Phenomenology of charm yukawa at the CEPC}
\subsection{Signal and backgrounds}
\begin{figure}[H]
\centering
\subfloat[]{\includegraphics[width=0.2\textwidth]{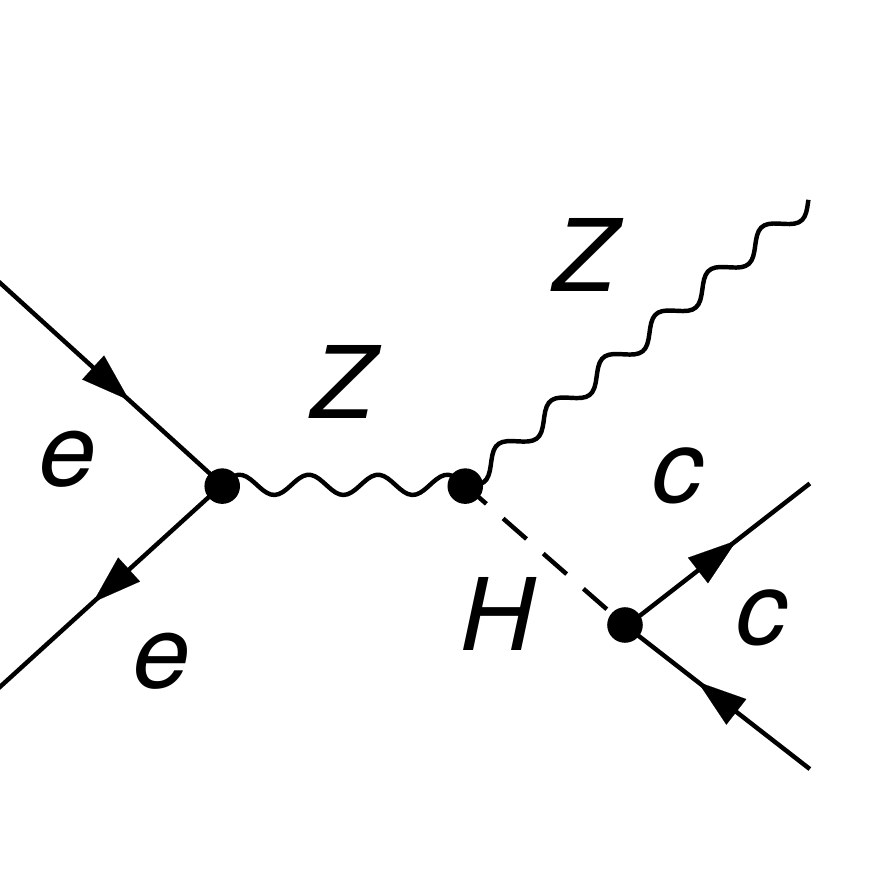}} \quad
\subfloat[]{\includegraphics[width=0.2\textwidth]{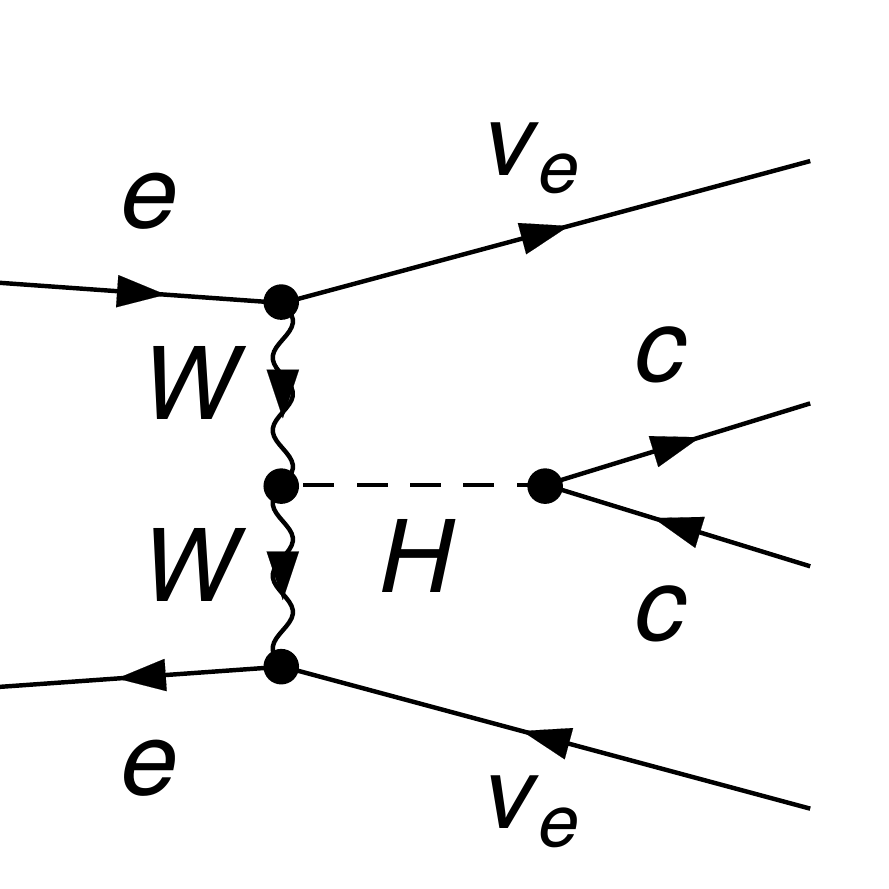}} \quad
\subfloat[]{\includegraphics[width=0.2\textwidth]{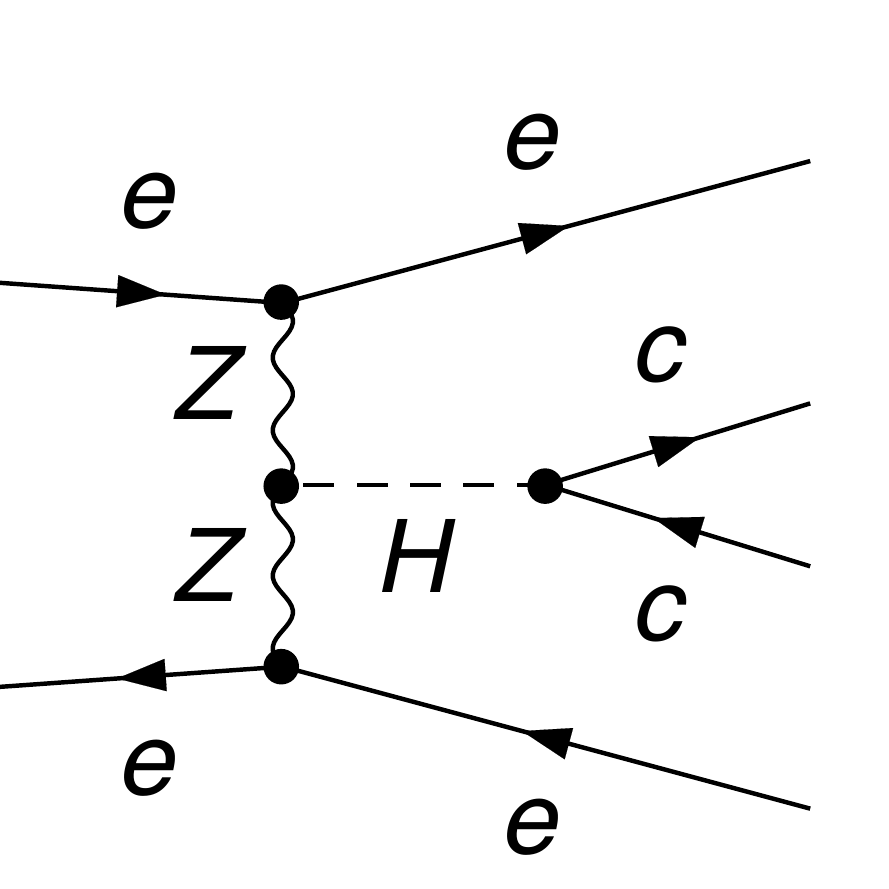}} \\
\subfloat[]{\includegraphics[width=0.2\textwidth]{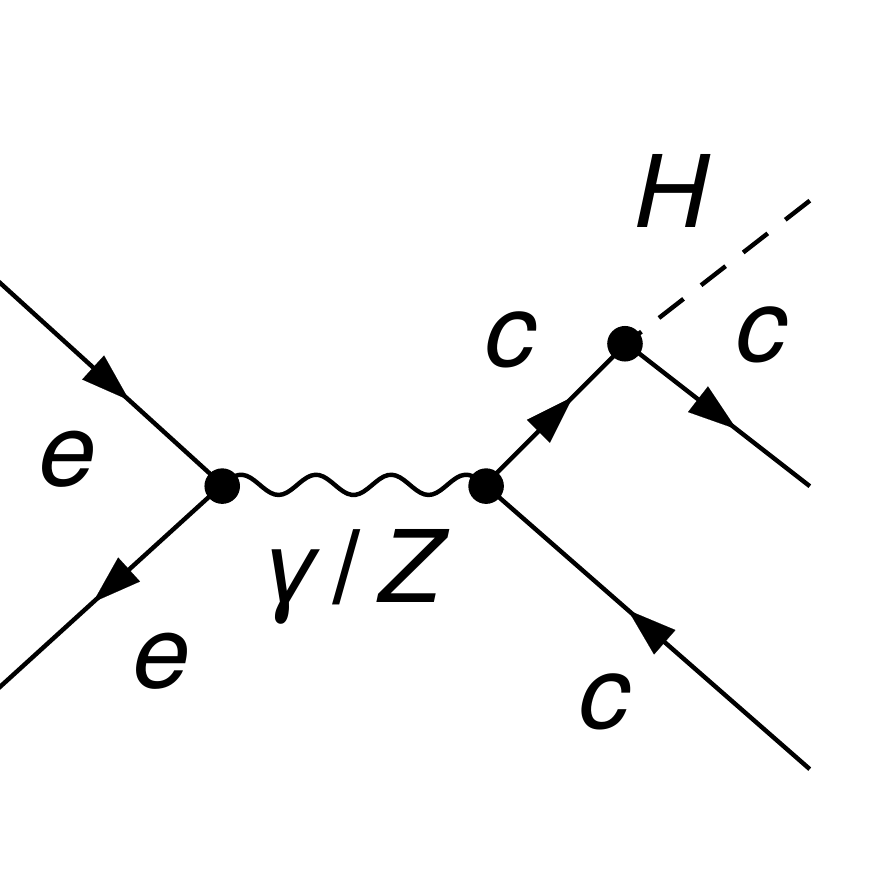}} \quad
\subfloat[]{\includegraphics[width=0.2\textwidth]{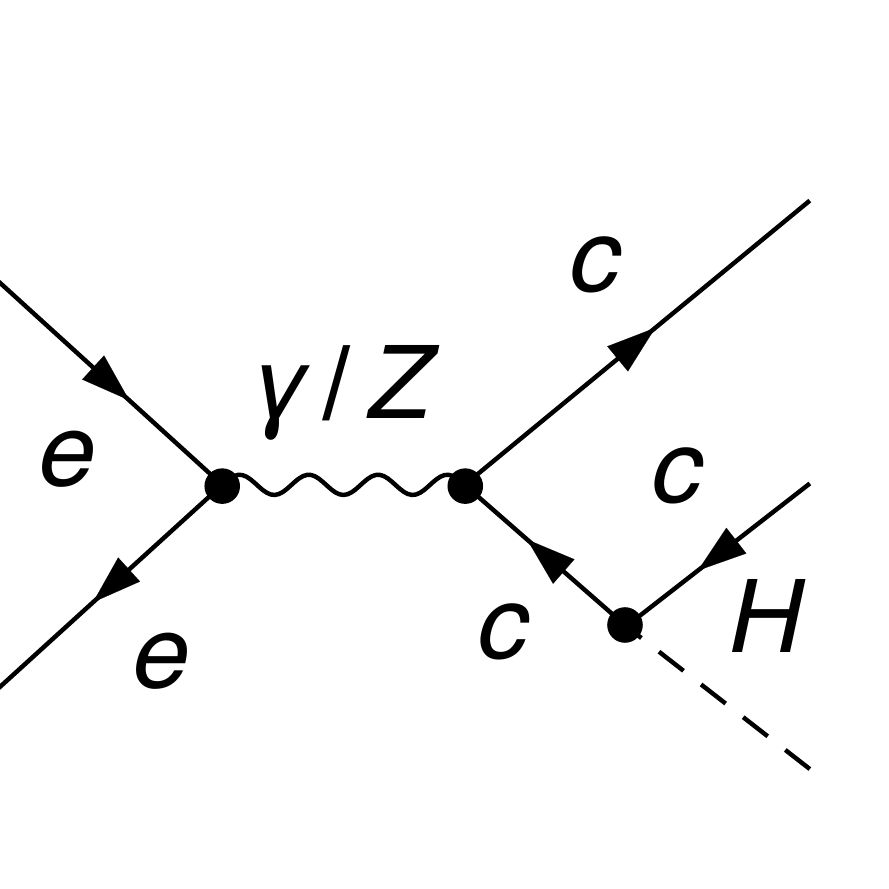}} \quad
\caption{Processes containing Higgs-charm Yukawa coupling at the CEPC.}
\label{process2}
\end{figure}

The CEPC is expected to make an excellent measurement of the charm Yukawa coupling and constrain $\delta y_{c}$ to $\sim2\%$ \cite{An:2018dwb}. In principle, the Higgs-charm Yukawa coupling could be probed in $e^{+}e^{-}$ collision through two modes: decay and direct production. Fig.\ref{process2}.~(a) shows the associated production of the Higgs and Z boson (ZH). (b) and (c) are vector-boson fusion (VBF) for charged current (CC) and neutral current (NC) processes respectively. The first three processes are used to probe Higgs-charm Yukawa coupling through Higgs boson decay to a charm quark pair directly. (d) and (e) are the processes of the positron and electron scattering to produce a charm quark pair in association with a single Higgs in the final state, which provides an independent measurement of charm Yukawa.~With the large center-of-mass energy $\sqrt{s} \sim 240~\text{GeV}$ at the CEPC , the ZH process is the dominant process. Hence we focus on
\bea
e^{+}e^{-} \to Zh,~h\to c\bar{c} \nonumber
\eea
as the signal for studying the physical potential of measuring the Higgs-charm Yukawa coupling at the CEPC. In order to distinguish the signal and backgrounds,~we classify the signal into four categories according to different decay channels of the $Z$ boson in the final state: hadronic decay, muonic decay, electronic decay, and invisible decay with neutrinos produced.~The Monte-Carlo events are generated by {\it MadGraph5\_v2.6.4}~\cite{Alwall:2014hca} at parton level.~{\it Pythia6.420} \cite{Sjostrand:2006za} and {\it Delphes3.3.0} \cite{deFavereau:2013fsa} are used for parton shower and detector simulations respectively.  Basic cuts are the same as in section \ref{LHeC}. 

\subsection{hadronic decay channel: $Z \to q q$}
Here $q$ denotes all quarks and anti-quarks ($u,d,s,c,b$) that $Z$ boson could decay to.  As stated in section \ref{LHeC}, we separate final jets into three categories: light-jet ($j$), c-jet ($c$) and b-jet ($b$) in both the signal and backgrounds. The main irreducible backgrounds come from processes with four jets in the final states
\bea
e^{+}e^{-} \to c c j j,~ b b j j,~j j j j,~\text{etc}. \nonumber
\eea 
There are many QCD radiative processes associated with jet production in the final state. Nonetheless, because of the larger phase-space, the dominant contribution is still from the processes $e^{+}e^{-} \to ZZ / Z\gamma^{*} / \gamma^{*}\gamma^{*}~\text{and}~W^{+}W^{-}$ followed by gauge boson hadronic decay.~There is a specific irreducible background $e^{+}e^{-} \to Zh, Z\to c\bar{c}, h\to b\bar{b}$, where the Higgs production process is the same as the signal, but the decay product is different. The kinematic feature of this process is identical to the signal and we could only expect  to suppress it with jet tagging. There are also reducible backgrounds. The dominant one is
 \bea
e^{+}e^{-} \to \ell \nu_{\ell} c j,~ \ell \nu_{\ell}  j j, \nonumber
\eea 
where $\ell=e,\mu$. These are pure EW/QED process, and the lepton and neutrino mainly come from the leptonic decay of one of the gauge bosons in the $ZZ, Z\gamma,W^{+}W^{-}$ pairs. The cross sections of the signal and backgrounds are given in Table.\ref{xsection1} with a 2.9\% branching ratio for Higgs decay to a charm pair.
\begin{table}[H]
\renewcommand\arraystretch{1.0}
\centering
\begin{tabular}{|c||c|c|c|c|}
\hline
\hline
S\&B & signal & four jets & $ccbb$~($Zh$) & $\ell \nu_{\ell} + \text{two jets}$  \\
\hline
cross-section & 4.65 & 4608.1 & 22.7 & 3704  \\
\hline
\hline
\end{tabular}
\caption{Cross-sections (in fb) for the signal, the irreducible background (four jets), $ccbb$~($Zh$) and the reducible background ($\ell \nu_{\ell} + \text{two jets}$) with $Br^{SM}(h\to c\bar{c})\approx 2.9\%$. }
\label{xsection1}
\end{table} 

\subsubsection*{selection cuts}
\begin{figure}[H]
\centering
\subfloat[The invariant mass $M(c,c)$ of two c-jets]{\includegraphics[width=0.4\textwidth]{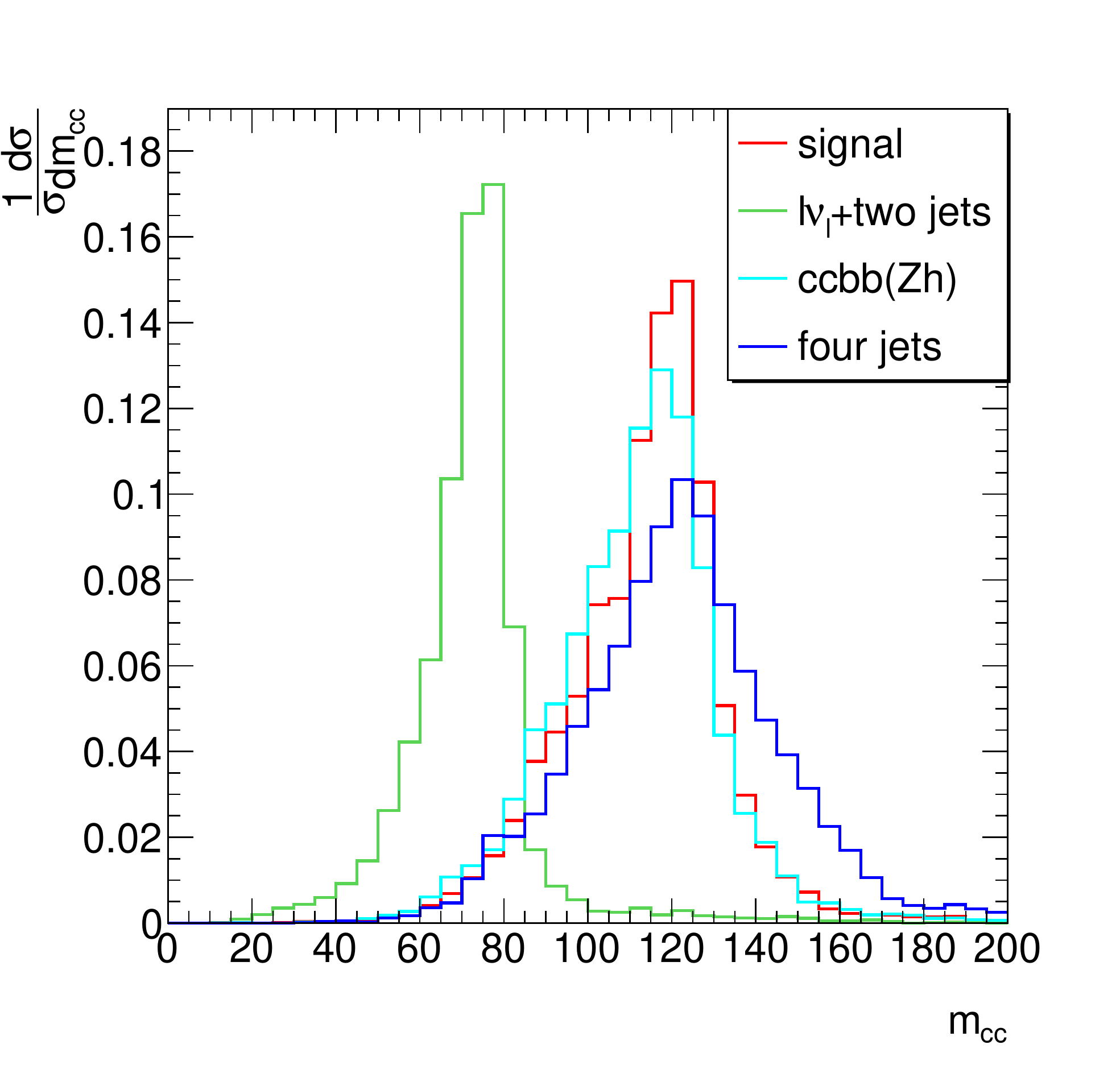}}
\subfloat[The invariant mass $M(j,j)$ of the other two jets]{\includegraphics[width=0.4\textwidth]{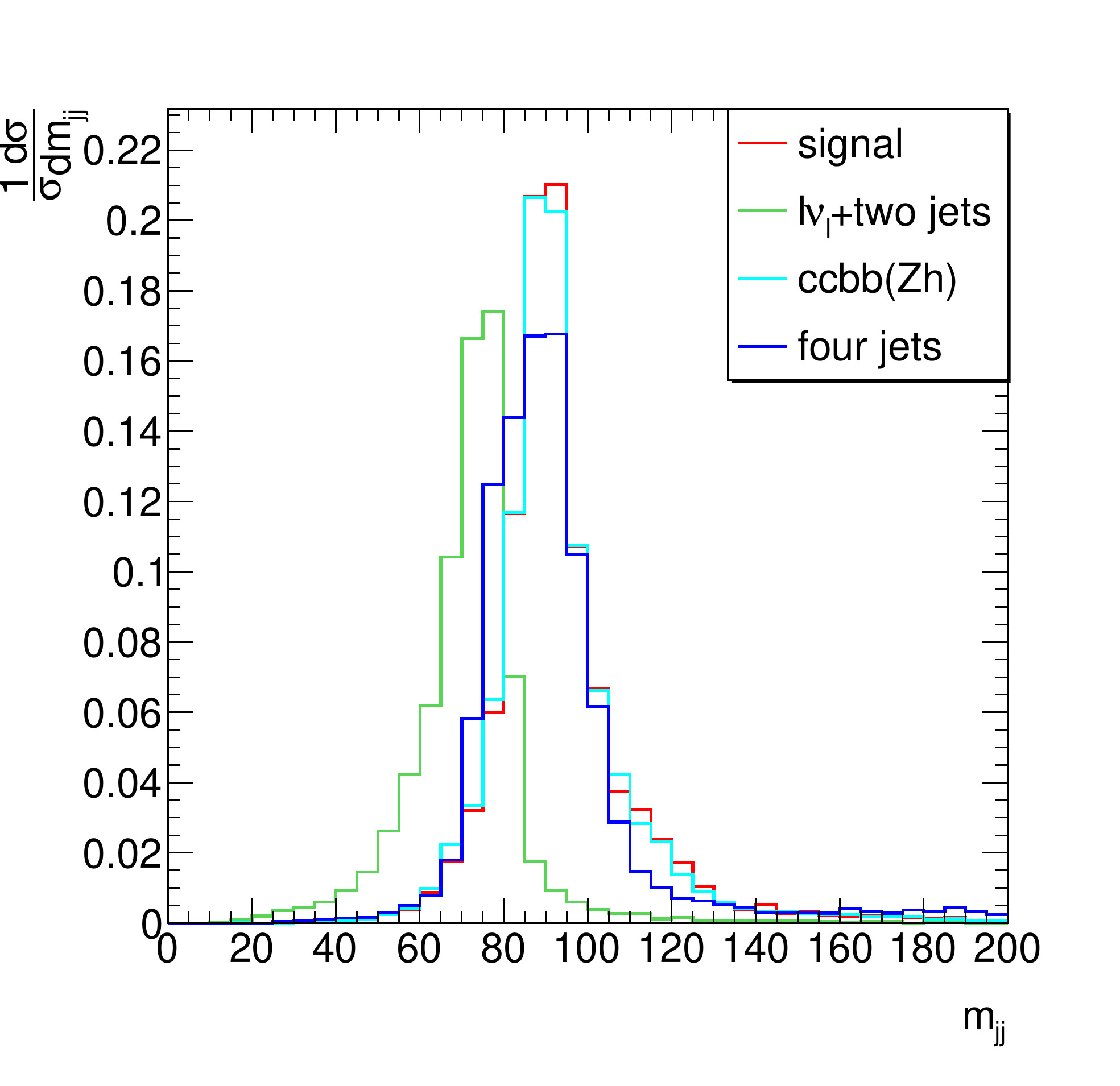}} 
\caption{The normalized distribution of the invariant mass $M(c,c)$ (left panel) and $M(j,j)$ (right panel) after full detector simulation.}
\label{mass1}
\end{figure}

In order to reduce the background overshadowing the signal, effective cuts are needed. As with the LHeC, the invariant mass $M(c,c)$ of the two c-jets might be a good kinematic observable to distinguish the signal and backgrounds. As the mother particle of a jet is unknown, we pick two jets whose invariant mass is closest to 125 GeV to reconstruct $M(c,c)$. The two chosen jets are considered to be c-jets. The other jets are then either from $Z$ boson decays for the signal, or from $W$ boson/QCD radiations in the background case.  The invariant mass $M(j,j)$ of the two remaining jets might be helpful for distinguishing signal and backgrounds as well. We plotted the distributions of $M(c,c)$ and $M(j,j)$ in Fig.\ref{mass1} after full detector simulation, where the red line corresponds to the signal, the green line corresponds to the reducible background, and the cyan and blue lines correspond to the above two irreducible backgrounds. We find an obvious deviation between the signal and $\ell \nu_{\ell}+\text{two jets}$ background since the jets in this reducible background dominantly come from W boson decays.~The invariant mass distribution of the $ccbb$~($Zh$) backgrounds has a large overlap with the signal as expected. {There is also a large overlap in $M(c,c)$ distribution between the signal and four jet background because two jets from decays of two different gauge bosons, chosen to reconstruct $M(c,c)$, just have an invariant mass near 125 GeV.} The detector resolution is partly responsible for the overlap as well. The peak in the $M(c,c)$ distribution has a deviation from Higgs mass due to the variation of final jet momenta during the hadronization and parton shower. Moreover, we require there are no common jets  in the reconstruction of $M(c,c)$ and $M(j.j)$ to better discriminate between the signal and backgrounds.~The selection cuts and cut-flow for the signal and background events are shown in Table.\ref{cut-flow1}. 

\begin{table}[H]
\renewcommand\arraystretch{0.7}
\centering
\begin{tabular}{|c||c|c|c|c|}
\hline
\hline
cuts & signal & four jets & $ccbb$~($Zh$) & $\ell \nu_{\ell} + \text{two jets}$  \\
\hline
\hline
basic cuts & 9300 & 9216200 & 45400 & 7408000 \\
\hline
$110~\rm{GeV}<$$M(c,c)$$<135~\rm{GeV}$ & 5188.47 & 4064730 & 22205.1 & 84451.2 \\
\hline
$|M(j,j)-91.18|<10~\rm{GeV}$ & 3170.37 & 2261990 & 13002.6 & 13334.4  \\
\hline
no common jets in $M(c,c)$ and $M(j.j)$ & 538.47 & 387516 & 1638.94 & 74.1 \\
\hline
$|\eta_{c,j}|<1$ in $M(c,c)$ and $M(j.j)$ & 333.87 & 151978 & 1021.5 & 59.26 \\
\hline
tagging i & 13.34 & 560.46 & 2.56 & 2.84 \\
tagging ii & 30.01 & 964.82 & 12.38 & 4.06 \\
tagging iii & 56.06 & 2121.25 & 63.89 & 8.33 \\
\hline
\hline
\end{tabular}
\caption{Cut-flow of the signal and background events at CEPC with $\mathcal{L}= 2~\text{ab}^{-1}$. The last row presents the number of events after kinematic cuts times the corresponding tagging efficiency given in Table.\ref{tag}. }
\label{cut-flow1}
\end{table} 

The cut-flow shows $M(c,c)$ and $M(j,j)$ selections eliminate about two thrids of the signal and irreducible backgrounds, while only $\mathcal{O}(6\times10^{-2})$ of the reducible background remains. The most efficient kinematic cut is the requirement of no common jets participating in the reconstruction of $M(c,c)$ and $M(j.j)$, which suppresses the two irreducible backgrounds by a factor $\sim 10$, and leads to a reducible background smaller than the signal. Note that  almost all $\ell \nu_{\ell} + \text{two jet}$ events only have two jets in the final state, in which case common jets are unavoidable when constructing $M(c,c)$ and $M(j.j)$. The cut of the pseudorapidity is used to restrict the Higgs and $Z$ boson to the central region and suppress some t-channel processes from the backgrounds. After accounting for the tagging efficiency given in Table.\ref{tag}, all backgrounds are reduced by approximately  two orders. $\ell \nu_{\ell} + \text{two jets}$ is essentially negligible.

\subsection{electronic decay channel: $Z \to e^{+}e^{-}$}
In this channel, Z boson decays to a positron and electron pair. As before, the jets from the backgrounds are divided into three categories according to their invariant mass. The main irreducible backgrounds come from processes with a $e^{+}e^{-}$ pair and two jets in the final states
\bea
e^{+}e^{-} \to e^{+} e^{-} c{c},~ e^{+} e^{-} j j,~e^{+} e^{-} b b. \nonumber
\eea
All final states are produced by pure EW/QED processes without QCD radiations. The dominant contribution comes from $e^{+}e^{-} \to ZZ / Z\gamma^{*} / \gamma^{*}\gamma^{*}~\text{and}~W^{+}W^{-}$ followed by gauge boson hadronic and leptonic decays.  Process from $Z, \gamma$ or $W$ bremsstrahlung through t-channel, e.g., $e^{+}e^{-} \to e^{+}e^{-}Z,~Z \to j j$, also need be considered, which features an event shape that is mostly in the forward or backward directions.  There is the $e^{+}e^{-} \to Zh, Z\to e^{+} e^{-}, h\to b\bar{b}$ process with the same kinematic features as the signal, which is added to the $e^{+} e^{-} b b$ background. We expect to reduce it with jet tagging. Both the signal and backgrounds in this channel have small rates as a result of the smallness of the coupling strength. The cross sections of the signal and backgrounds are shown in Table.\ref{xsection2}
\begin{table}[H]
\renewcommand\arraystretch{1.0}
\centering
\begin{tabular}{|c||c|c|c|c|}
\hline
\hline
S\&B & signal & $e^{+} e^{-} c {c}$ & $e^{+} e^{-} j j$ & $e^{+} e^{-} b {b}$   \\
\hline
cross-section & 0.232 & 31.16 & 92.5 & 35.85   \\
\hline
\hline
\end{tabular}
\caption{Cross-sections (in fb) for the signal, and three irreducible backgrounds $e^{+} e^{-} c {c}$,~$e^{+} e^{-}jj$ and $e^{+} e^{-} bb$ with $Br^{SM}(h\to c\bar{c})\approx 2.9\%$. }
\label{xsection2}
\end{table} 

\subsubsection*{selection cuts}
\begin{figure}[H]
\centering
\subfloat[The invariant mass $M(c,c)$ of two c-jets]{\includegraphics[width=0.4\textwidth]{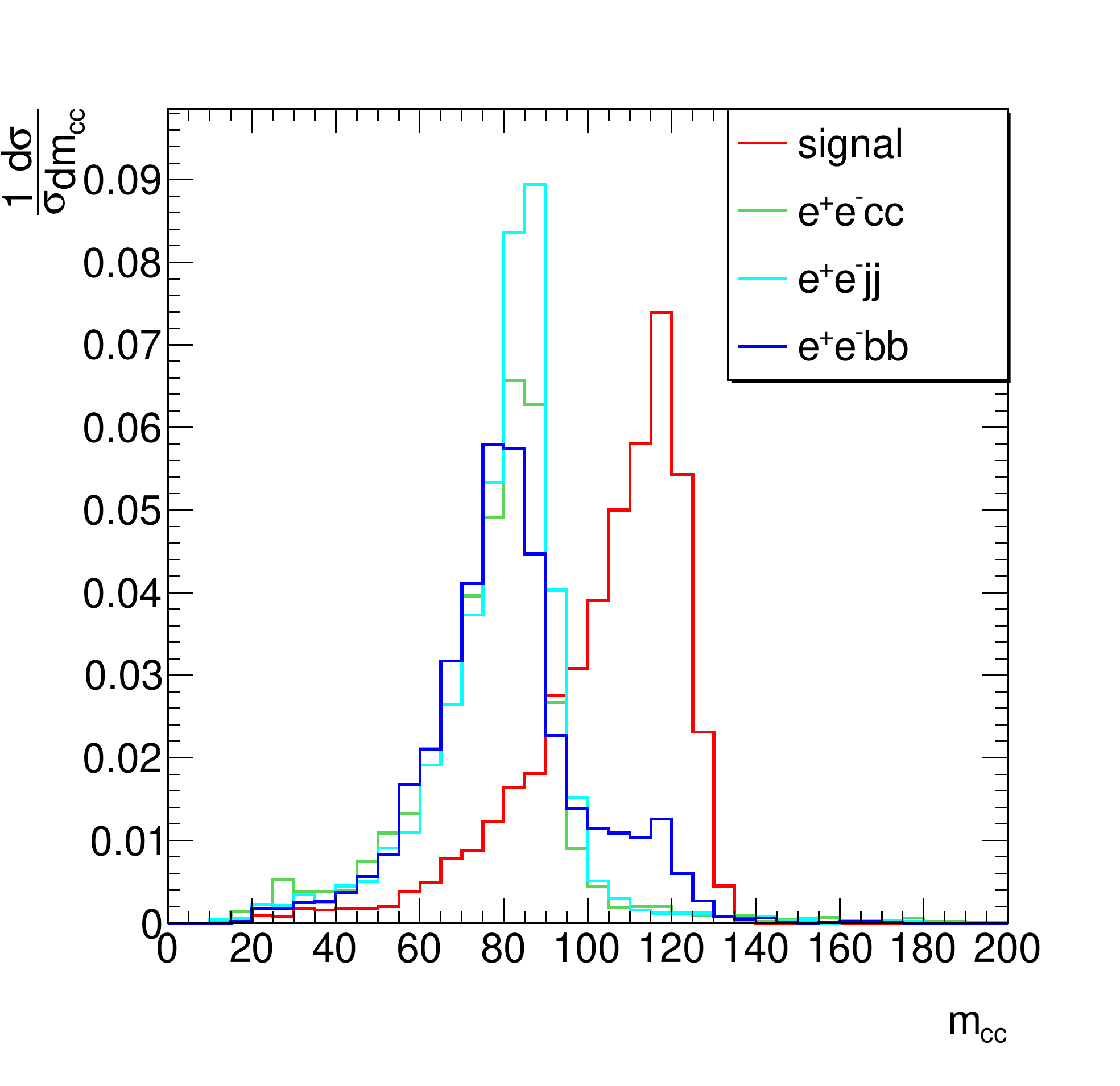}}
\subfloat[The invariant mass $M(e^{+},e^{-})$ of the electron pair]{\includegraphics[width=0.4\textwidth]{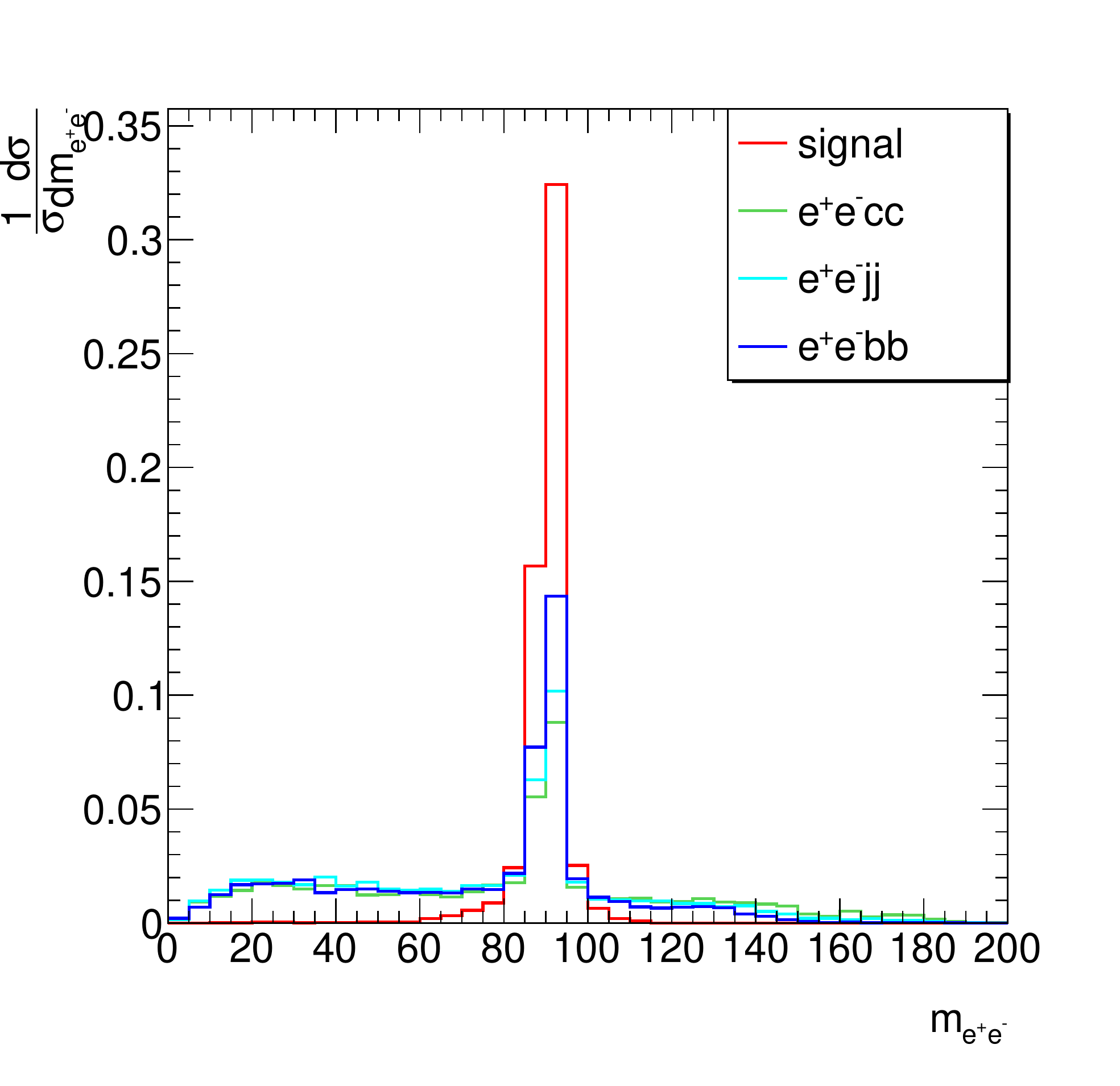}} 
\caption{The normalized distribution of the invariant mass $M(c,c)$ (left panel) and $M(e^{+},e^{-})$ (right panel) after full detector simulation.}
\label{mass2}
\end{figure}

$M(c,c)$ of the two c-jets and $M(e^{+},e^{-})$ of the positron-electron pair are important observables for distinguishing the signal and backgrounds, as shown in Fig.\ref{mass2}. The red line corresponds to the signal, the green, cyan and blue lines correspond to the three backgrounds respectively. Note the $M(c,c)$ peak for the signal in Fig.\ref{mass1} is broader than in Fig.\ref{mass2} since it is more likely to misidentify a c jet when constructing $M(c,c)$ in the case of hadronic $Z$ decays. As before, the  $M(c,c)$ peak still has a small deviation from 125 GeV because of effects of the hadronization and parton shower  that lead to the variation of final jet momenta. Yet the distributions of $M(e^{+},e^{-})$ have  sharp peaks at the $Z$ boson mass ($m_{Z}=91.18~\text{GeV}$), thanks to the precise identification of the lepton final states. The selection cuts and cut-flow for the signal and background events are shown in Table.\ref{cut-flow2} 

\begin{table}[H]
\renewcommand\arraystretch{0.7}
\centering
\begin{tabular}{|c||c|c|c|c|}
\hline
\hline
cuts & signal & $e^{+} e^{-} c {c}$ & $e^{+} e^{-} j j$ & $e^{+} e^{-} b {b}$  \\
\hline
\hline
basic cuts & 464 & 62320 & 185000 & 71700 \\
\hline
$110~\rm{GeV}<$$M(c,c)$$<135~\rm{GeV}$ & 176.39 & 853.86 & 2057.08 & 4267.86 \\
\hline
$|M(e^+e^-))-91.18|<10~\rm{GeV}$ & 167.15 & 353.74 & 1097.11 & 3729.45  \\
\hline
all $e^{+},e^{-}$ with $|\eta_e|<2$ & 152.14 & 219.57 & 891.4 & 3401.15 \\
\hline
tagging i & 6.09& 8.78 & 0.971 & 8.5 \\
tagging ii & 13.69 & 19.76 & 0.971 & 41.15 \\
tagging iii & 25.57 & 36.91 & 2.23 & 212.57 \\
\hline
\hline
\end{tabular}
\caption{Cut-flow of the signal and background events at CEPC with $\mathcal{L}= 2~\text{ab}^{-1}$. The last row presents the number of events after kinematic cuts times the corresponding tagging efficiency given in Table.\ref{tag}. }
\label{cut-flow2}
\end{table} 

$M(c,c)$ and $M(e^{+},e^{-})$ cuts reduce all the backgrounds by approximately one to two orders. The Pseudorapidity cut is used to eliminate events with the forward positron and electron in the final state, but the improvement is marginal since many of these events are already excluded by the invariant mass cuts. The remaining  $e^{+} e^{-} b {b}$ events are mostly from the associated production of the Higgs and Z boson, and are difficult to be distinguished from the signal kinematically. Fortunately, $b$ tagging brings it down by a factor of 5 after the kinematic cuts.

\subsection{muonic decay channel: $Z \to \mu^{+}\mu^{-}$}
The event selection in muon channel is similar to the electron channel case. The main irreducible backgrounds are the following processes:
\bea
e^{+}e^{-} \to \mu^{+} \mu^{-} c{c},~ \mu^{+} \mu^{-} j j,~\mu^{+} \mu^{-} b b. \nonumber
\eea
These are also pure EW/QED processes. A difference between  muon and  electron channels is that the gauge boson bremsstrahlung processes through t-channel disappear, since there is no flavor changing the neutral current in the SM. Of course, the s-channel process still contributes. $e^{+}e^{-} \to Zh, Z\to \mu^{+} \mu^{-}, h\to b\bar{b}$ is part of the $\mu^{+} \mu^{-} b b$ background as with the electron channel. The cross sections of the signal and backgrounds are shown in Table.\ref{xsection3}, which are still small. But without the t-channel gauge boson bremsstrahlung in background, the significance and signal-to-background would be improved.
\begin{table}[H]
\renewcommand\arraystretch{1.0}
\centering
\begin{tabular}{|c||c|c|c|c|}
\hline
\hline
S\&B & signal & $\mu^{+} \mu^{-} c {c}$ & $\mu^{+} \mu^{-} j j$ & $\mu^{+} \mu^{-} b {b}$   \\
\hline
cross-section & 0.232 & 10.76 & 36.47 & 18.16   \\
\hline
\hline
\end{tabular}
\caption{Cross-sections (in fb) for the signal, and three irreducible backgrounds $\mu^{+} \mu^{-} c {c}$,~$\mu^{+} \mu^{-} jj$ and $\mu^{+} \mu^{-} bb$ with $Br^{SM}(h\to c\bar{c})\approx 2.9\%$. }
\label{xsection3}
\end{table} 

\subsubsection*{selection cuts}
\begin{figure}[H]
\centering
\subfloat[The invariant mass $M(c,c)$ of two c-jets]{\includegraphics[width=0.4\textwidth]{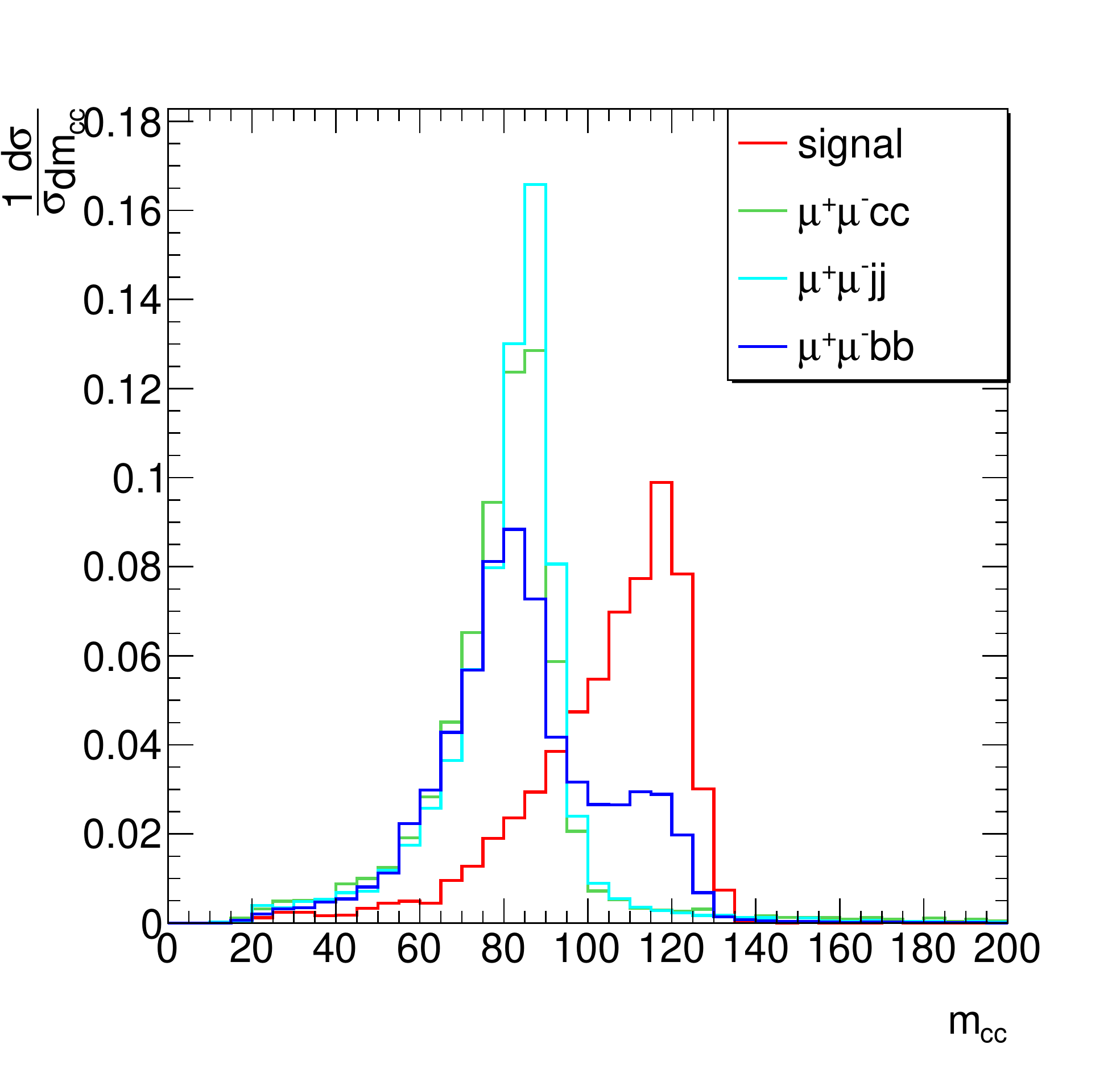}}
\subfloat[The invariant mass $M(\mu^{+},\mu^{-})$ of the muon pair]{\includegraphics[width=0.4\textwidth]{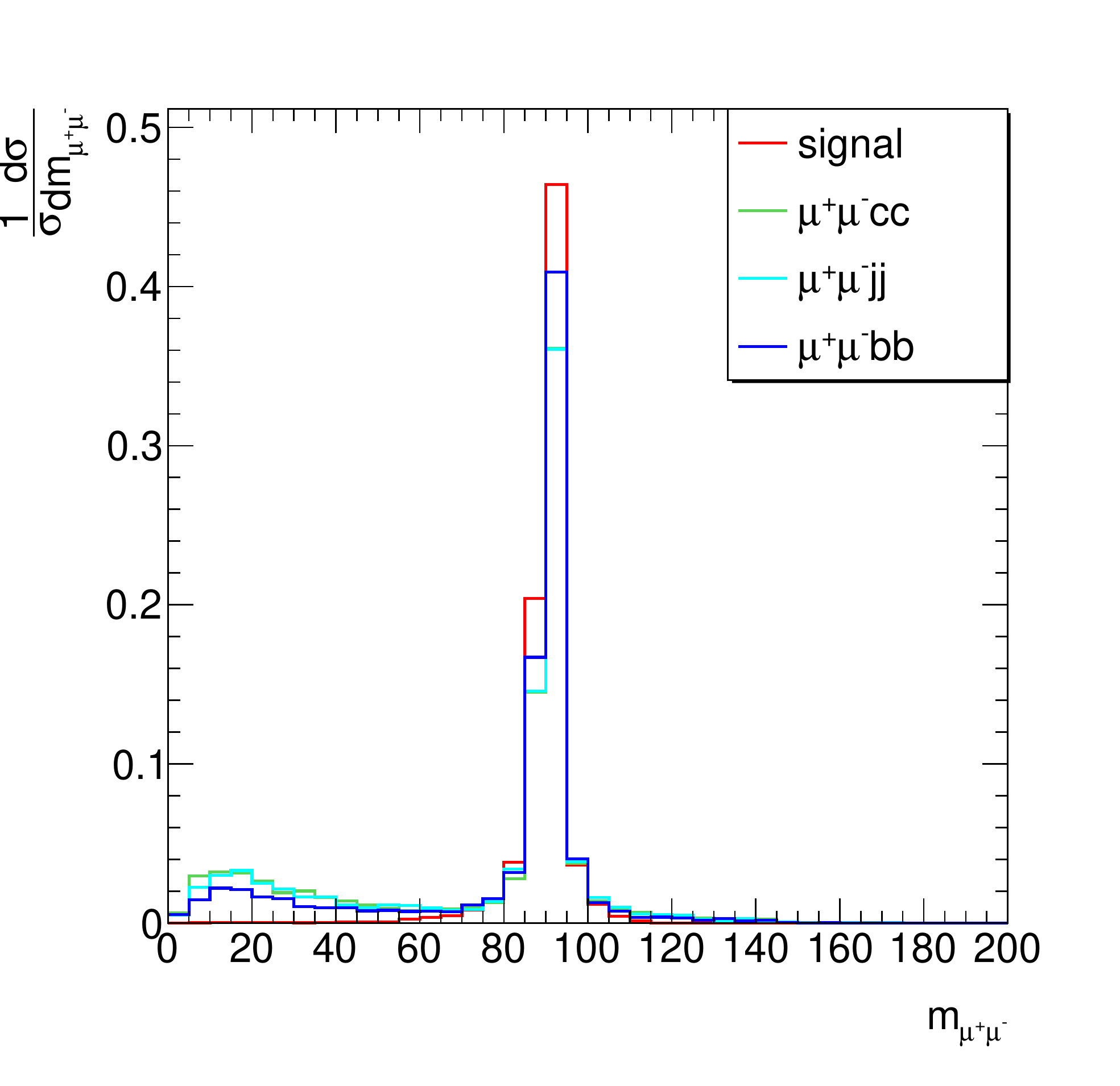}} 
\caption{The normalized distribution of the invariant mass $M(c,c)$ (left panel) and $M(e^{+},e^{-})$ (right panel) after full detector simulation.}
\label{mass3}
\end{figure}

We plot the invariant mass $M(c,c)$ of the two c-jets and $M(\mu^{+},\mu^{-})$ of a muon pair in Fig.\ref{mass3}.  The red line corresponds to the signal, the green, cyan and blue lines correspond to the three different backgrounds respectively. The signal $M(\mu^{+},\mu^{-})$ has a sharp peak at the Z boson mass ($m_{Z}=91.18~\text{GeV}$), while the backgrounds have a flat tail in the small $M(\mu^{+},\mu^{-})$ region, where $\gamma^{*}\rightarrow \mu^{+}\mu^{-}$ dominates. Hence the cut near the 125 GeV is effective for reducing the backgrounds. Moreover, the pseudorapidity cut is no longer needed due to lack of the forward final states from the t-channel processes. In a word, only two kinematic cuts are set in Table.\ref{cut-flow3}.

\begin{table}[H]
\renewcommand\arraystretch{0.7}
\centering
\begin{tabular}{|c||c|c|c|c|}
\hline
\hline
cuts & signal & $\mu^{+} \mu^{-} c {c}$ & $\mu^{+} \mu^{-} j j$ & $\mu^{+} \mu^{-} b {b}$  \\
\hline
\hline
basic cuts & 464 & 21520 & 72940 & 36320 \\
\hline
$110~\rm{GeV}<$$M(c,c)$$<135~\rm{GeV}$ & 170.25 & 335.42 & 1008.46 & 3585.93 \\
\hline
$|M(\mu^+\mu^-)-91.18|<10~\rm{GeV}$ & 158.48 & 189.58 & 586.89 & 3266.35  \\
\hline
tagging i & 6.34 & 7.85 & 0.639 & 8.16 \\
tagging ii & 14.26 & 17.06 & 0.639 & 39.52 \\
tagging iii & 26.36 & 31.87 & 1.47 & 204.147 \\
\hline
\hline
\end{tabular}
\caption{Cut-flow of the signal and background events at CEPC with $\mathcal{L}= 2~\text{ab}^{-1}$. \ref{tag}.The last row presents the number of events after kinematic cuts times the corresponding tagging efficiency given in Table.\ref{tag}. }
\label{cut-flow3}
\end{table} 

\subsection{invisible decay channel: $Z \to \nu_{\ell}\bar{\nu}_{\ell}$}
Cross sections in this channel increase prominently because of the large branching fraction $Br(Z\to \nu_{\ell}\bar{\nu}_{\ell} ) \approx 20.5\%$. With no  QCD interactions, the main irreducible backgrounds are 
\bea
e^{+}e^{-} \to \nu_{\ell}\bar{\nu}_{\ell}  c{c},~ \nu_{\ell}\bar{\nu}_{\ell}  j j,~\nu_{\ell}\bar{\nu}_{\ell} b b, \nonumber
\eea
where $\ell=e,\mu,\tau$.~The dominant subprocess chains are $e^{+}e^{-} \to ZZ, Z \to \text{hadrons}, Z \to \nu_{\ell} \bar{\nu}_{\ell}$ and $e^{+}e^{-} \to Z\gamma^{*}, \gamma^{*} \to \text{hadrons}, Z \to \nu_{\ell} \bar{\nu}_{\ell}$. The t-channel gauge boson bremsstrahlung through charged current processes appears as a background in this channel, but only for the case where $\nu_{\ell} = \nu_{e}$. The cross sections of the signal and backgrounds are given in Table.\ref{xsection4}.
\begin{table}[H]
\renewcommand\arraystretch{1.0}
\centering
\begin{tabular}{|c||c|c|c|c|}
\hline
\hline
S\&B & signal & $\nu_{\ell}\bar{\nu}_{\ell} c {c}$ & $\nu_{\ell}\bar{\nu}_{\ell} j j$ & $\nu_{\ell}\bar{\nu}_{\ell} b {b}$   \\
\hline
cross-section & 1.37 & 52.08 & 179.1 & 104.2   \\
\hline
\hline
\end{tabular}
\caption{Cross-sections (in fb) for the signal, and three irreducible backgrounds $\nu_{\ell}\bar{\nu}_{\ell} c{c}$,~$\nu_{\ell}\bar{\nu}_{\ell} j {j}$ and $\nu_{\ell}\bar{\nu}_{\ell} bb$ with $Br^{SM}(h\to c\bar{c})\approx 2.9\%$. }
\label{xsection4}
\end{table}

\subsubsection*{selection cuts}
\begin{figure}[H]
\centering
\subfloat[The transverse momentum $\slashed{E}_{T}$]{\includegraphics[width=0.4\textwidth]{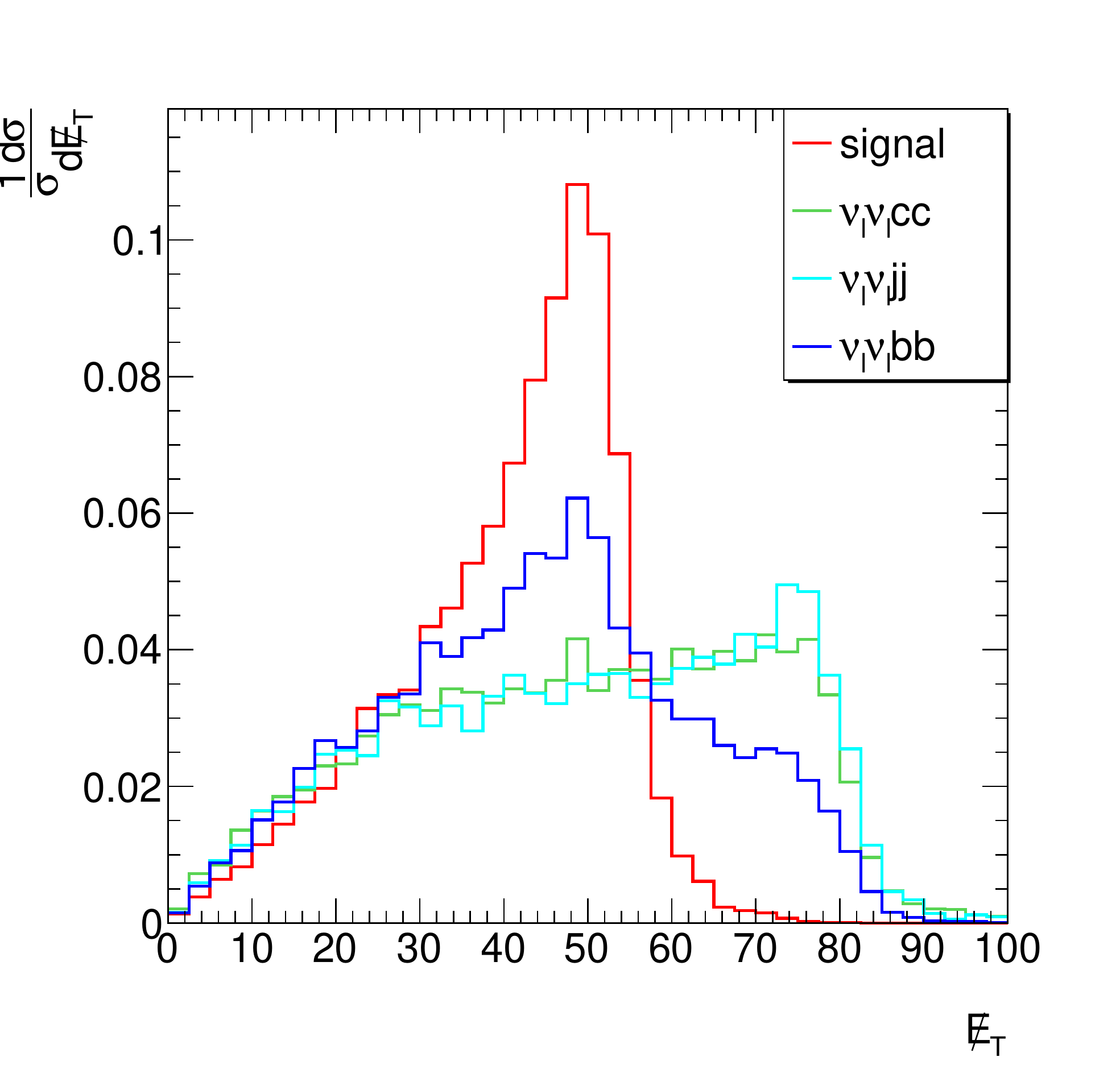}}
\subfloat[The transverse mass $MT$]{\includegraphics[width=0.4\textwidth]{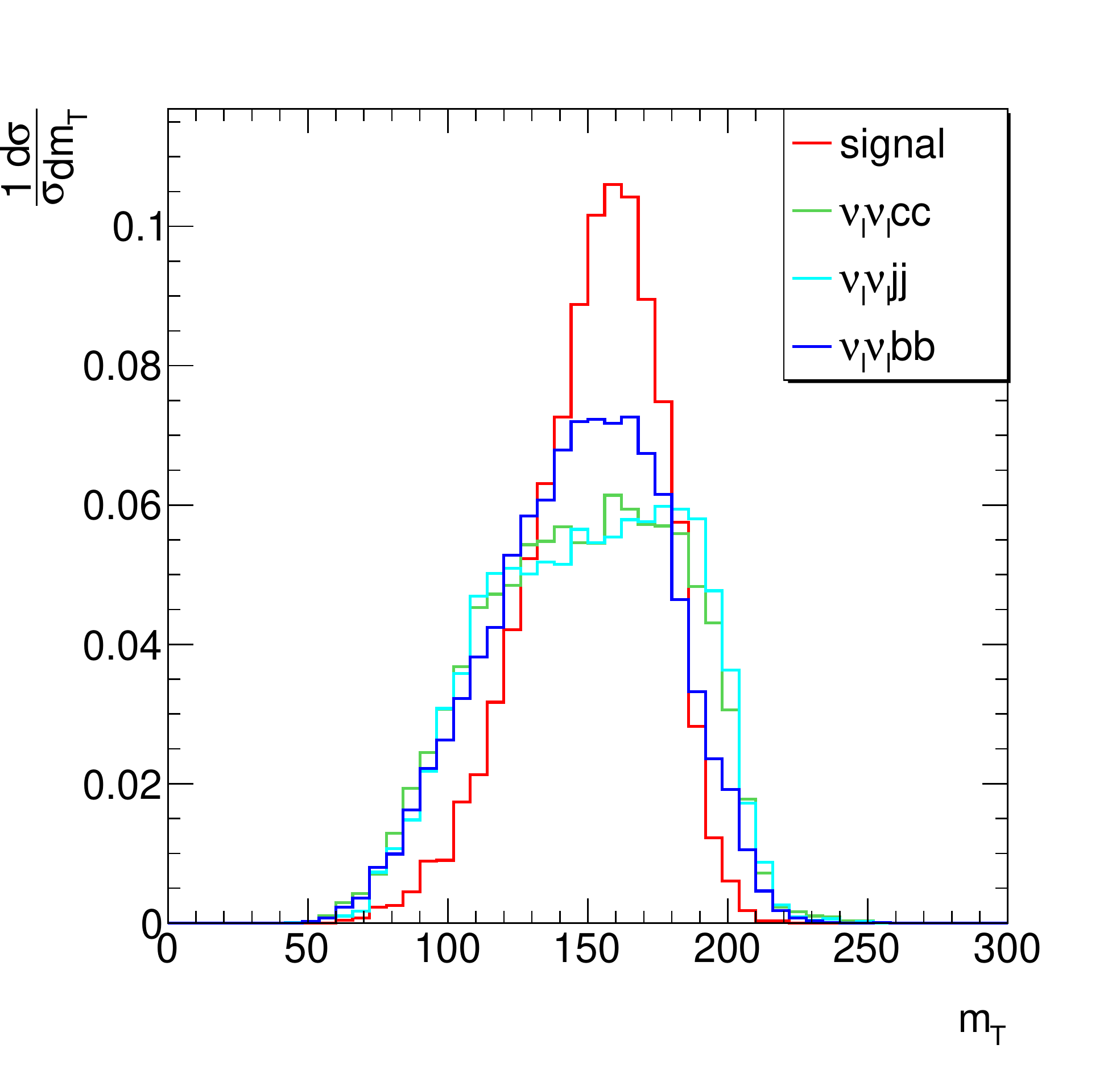}}  \\
	\subfloat[The invariant mass $M(c,c)$ of two c-jets]{\includegraphics[width=0.4\textwidth]{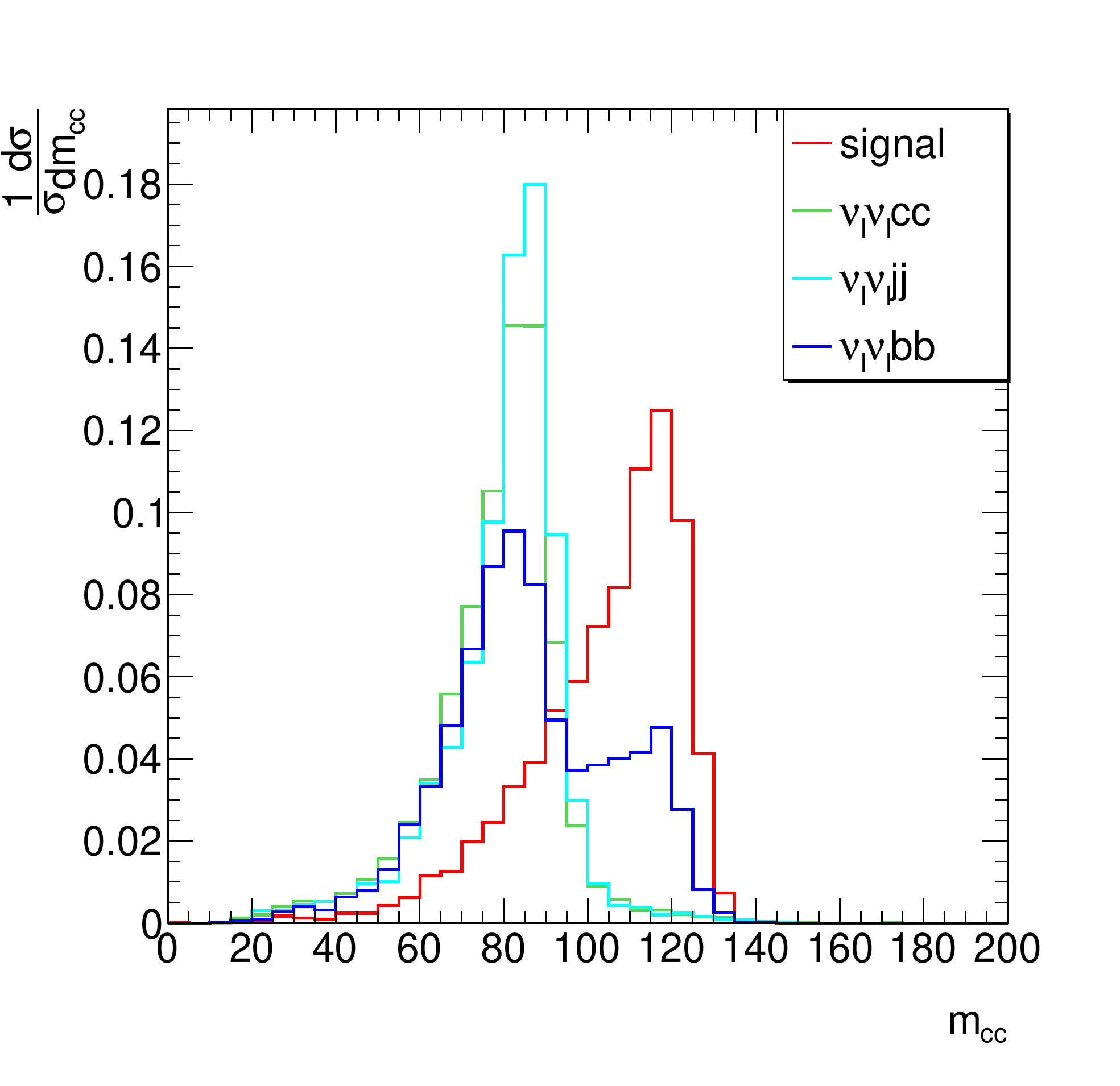}}
\caption{The normalized distribution of the missing transverse momentum $\slashed{E}_{T}$ (top left panel), the transverse mass $MT$ (top right panel), and the invariant mass $M(c,c)$ (bottom panel) after full detector simulation.}
\label{mass4}
\end{figure}

We find the signal cross section 50 times larger than those in previous leptonic channels. However, the appearance of invisible particles is a challenge in constructing kinematic observables in the final state, since we can only obtain their transverse momenta by momentum conservation. That is to say, one can not reconstruct the invariant mass of the two invisible neutrinos as the full information of their momenta is unavailable. Therefore, instead of the invariant mass $M(\nu_{\ell},\bar{\nu}_{\ell})$, we appeal to the distributions of the missing transverse momentum, and of the transverse mass $MT \equiv \sqrt{\slashed{E}_{T}^{2}+M^{2}(c,c)}+\slashed{E}_{T}$ of the $c\bar{c}$ pair, as shown in Fig.\ref{mass4}. The red line corresponds to the signal, the green, cyan and blue lines correspond to the three different backgrounds respectively.~The $\slashed{E}_{T}$ of the signal tends to be smaller compared to the background because the large $M(c,c)\sim 125$~GeV leads to a relatively small momentum of the recoil system. However, as $M(c,c)$ and $\slashed{E}_{T}$ are inversly correlated, the $MT$ distributions are less sensitive to the variation of $M(c,c)$ in various processes. The $M(c,c)$ distributions are also plotted in Fig.\ref{mass4} where the distinction between the signal and background is clearly shown as before. One can see that there is a bump from the $\mu^{+} \mu^{-} b {b}$ background near the peak region of the signal, which is from the contribution of the $e^{+}e^{-} \to Zh, Z \to \nu_{\ell}\bar{\nu}_{\ell}, h \to b \bar{b}$ subprocess. It will be reduced with jet tagging. The cut flow for this channel is shown in Table.\ref{cut-flow4}

\begin{table}[H]
\renewcommand\arraystretch{0.7}
\centering
\begin{tabular}{|c||c|c|c|c|}
\hline
\hline
cuts & signal & $\nu_{\ell}\bar{\nu}_{\ell} c {c}$ & $\nu_{\ell}\bar{\nu}_{\ell} j j$ & $\nu_{\ell}\bar{\nu}_{\ell} b {b}$  \\
\hline
\hline
basic cuts & 2740 & 104160 & 358200 & 208400 \\
\hline
the missing transverse momentum $\slashed{E}_{T}<55~\rm{GeV}$ & 2530.66 & 59319.1 & 197440 & 148318 \\
\hline
$110~\rm{GeV}<$$M(c,c)$$<135~\rm{GeV}$ & 951.6 & 1062.43 & 3402.9 & 23445 \\
\hline
$MT>130~\rm{GeV}$ & 927.76 & 958.27 & 3080.52 & 22569.7  \\
\hline
tagging i & 37.11 & 38.33 & 3.35 & 56.42 \\
tagging ii & 83.50 & 86.24 & 3.35 & 272.09 \\
tagging iii & 156.0 & 161.08 & 7.70 & 1410.61 \\
\hline
\hline
\end{tabular}
\caption{Cut-flow of the signal and background events at CEPC with $\mathcal{L}= 2~\text{ab}^{-1}$. The last row presents the number of events after kinematic cuts times the corresponding tagging efficiency given in Table.\ref{tag}.  }
\label{cut-flow4}
\end{table} 

The transverse momentum $\slashed{E}_{T}$ could reduce the $\mu^{+} \mu^{-} c {c}$ and $\mu^{+} \mu^{-} j j$ backgrounds by a factor of 2, while keeping approximately 92\% of the signal. The invariant mass $M(c,c)$ is still a useful quantity to suppress the backgrounds. In contrast, the transverse mass $MT$ is inefficient. After tagging all the jets, the remaining reducible background events are small and comparable to the signal. The greatest advantage of the invisible decay channel is the relatively large number of the signal events after all cuts, which is 6 times that in leptonic channel, and 3 times that in hadronic channel.

\subsection{Combination and Results}
\begin{figure}[H]
\centering
\subfloat[$Z$]{\includegraphics[width=0.45\textwidth]{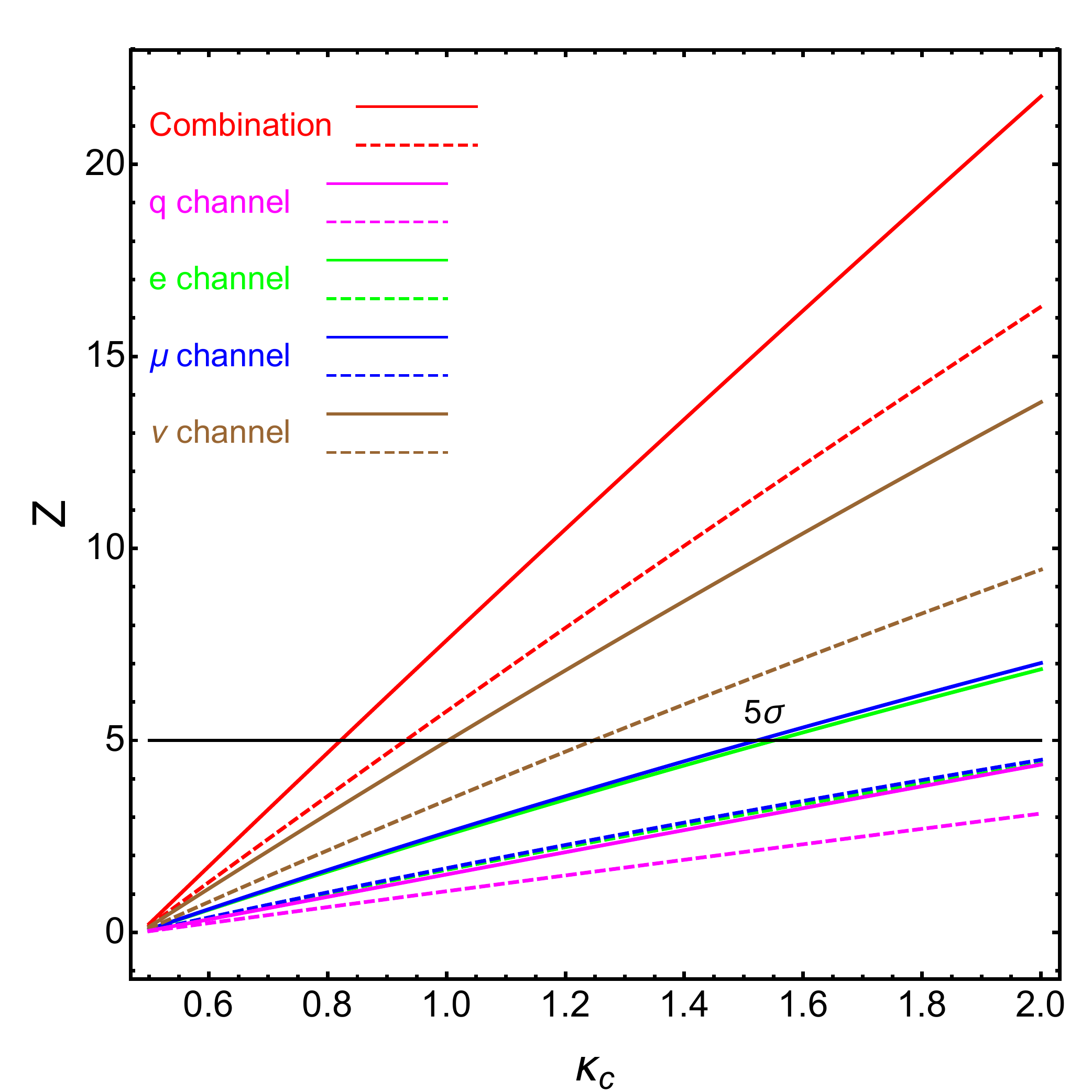}} \quad
\subfloat[$S/B$]{\includegraphics[width=0.45\textwidth]{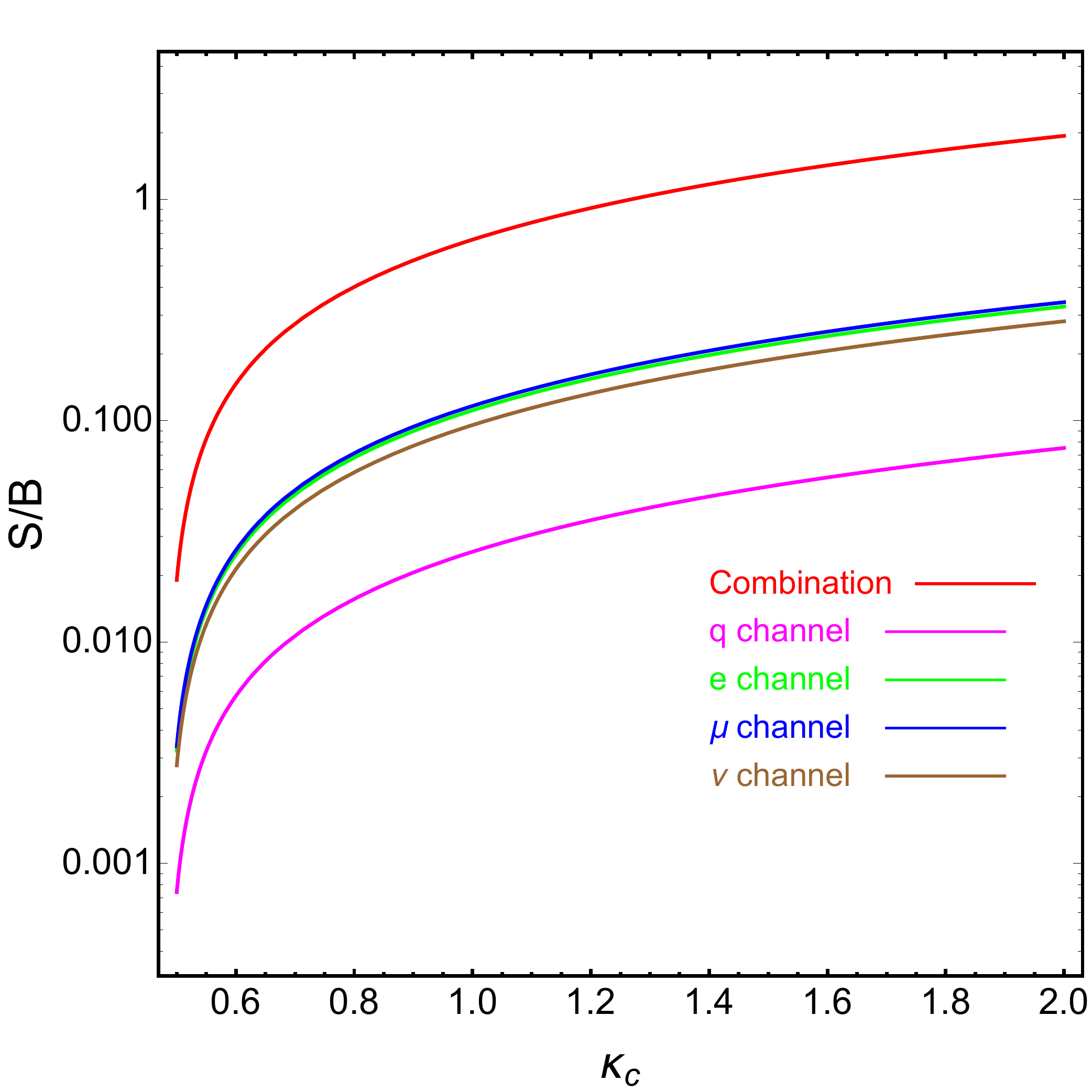}}
\caption{(a): the significance $Z$ varying with $\kappa_{c}$ at the CEPC; (b): the signal-to-background ($S/B$) varying with $\kappa_{c}$ at the CEPC. The center-of-mass energy of $\sqrt{s} \sim 240~\text{GeV}$. The tagging iii is used.}
\label{significance1}
\end{figure}

We compute the signal significance ($Z$) and the signal-to-background ($S/B$) in the same way as described in section~\ref{LHeC_results}. Here we also include the 1\% systematic error. $S$ and $B$ correspond to different signals and backgrounds in various channels. Using the Bessel formula to estimate the corresponding errors \cite{pdg2018}, we combine the four channels. The significance ($Z$) and the signal-to-background ($S/B$) are shown in Fig.\ref{significance1}, which are computed assuming the tagging efficiencies from the last row of Table.\ref{tag}. The magenta, green, blue and brown lines correspond to the hadronic, electronic, muonic, and invisible decay channels respectively. The red line is the combined result of the four channels. Solid and dashed lines represent different integrated luminosities. It is clear that the invisible decay channel shows a higher significance than others, while the signal-to-background for various channels shows less difference except for the hadronic decay, whose $S/B$ is much smaller. {When $\kappa_{c}=1$, the significance of the invisible decay could reach up to $3.4(5.0)\sigma$ with the 2(5) ab$^{-1}$ integrated luminosity. In contrast, the muonic can not reach $5\sigma$ until $\kappa_{c} \sim 1.5$ with the 5 ab$^{-1}$. } The hadronic is a difficult channel, which gets to $5\sigma$ at a much larger $\kappa_{c}$. After combining all channels, the significance goes up to $5.8(8.0)\sigma$ at $\kappa_{c} = 1$ when $\mathcal{L}=2(5)~ab^{-1}$ and the signal-to-background is close to 66\%.~Of course our rough estimation based on the MC simulation leaves out many reducible backgrounds in realistic detectors. We expect real data from {the detector} in the future will help improve the accuracy of the results. {More systematic studies are also needed to better constrain the precision of the charm Yukawa coupling determination \cite{An:2018dwb}.}

\section{\label{con}Conclusion}
Searching for the Higgs-charm Yukawa coupling can provide a crucial validation for the Higgs mechanism with the second generation charged fermions. It has been proven difficult to measure it at hadron colliders like the LHC. In this paper, we discuss physics potential for measuring the charm Yukawa at the LHeC and CEPC via WBF and ZH processes respectively, followed by the $h\to c\bar{c}$ inclusive decay. {Through simulation, we found that with a 60 GeV and -80\% polarization electron beam, and a 3 ab$^{-1}$ integrated luminosity, {the signal significance is $2\sigma$ (95\% C.L.) for $\kappa_{c} \simeq 1.18$ at the LHeC}, while at the CEPC, the signal significance goes up to $5.8(8.0)\sigma$ for the SM charm Yukawa with a 2(5) ab$^{-1}$ integrated luminosity after combination of four channels.} $e^{+}e^{-} \to Zh, Z \to \nu_{\ell}\bar{\nu}_{\ell}, h \to c \bar{c}$ shows a promising potential for measuring the charm Yukawa directly at positron-electron colliders because of the clean background and high production rate, even though there are invisible neutrinos in the final states. We expect complementary studies with more effective selection criteria and realistic detector level analysis to be helpful for probing the charm Yukawa accurately. 

\section*{Acknowledgments}
{We thank Dr. Tao Xu for his enthusiastic discussion and help.~We also thank one of the referees and the editor for their very useful suggestion.}~This work is supported by the National Science Foundation of China (11875232).

\bibliographystyle{utphysmcite}
\bibliography{hcc}

\end{document}